\documentclass{aa}
\usepackage{txfonts}
\usepackage{fullpage,graphicx}
\usepackage{setspace}
\usepackage{subfigure}
\usepackage{natbib}

\begin{document}

\title{The physical and dynamical structure of Serpens} 

\subtitle{Two very different sub-(proto)clusters} 

\author{A. Duarte-Cabral\inst{1}\thanks{Funded by the Funda{\c{c}}{\~a}o 
  para a Ci{\^e}ncia e a Tecnologia (Portugal)}  
  \and G.~A. Fuller\inst{1} 
  \and N. Peretto\inst{1}
  \and J. Hatchell\inst{2}
  \and E.~F. Ladd\inst{3}
  \and J. Buckle \inst{4,5} 
  \and J. Richer \inst{4,5}
  \and S.~F. Graves \inst{4,5}
  } 

\offprints{Ana Duarte Cabral, \email{Ana.Cabral@postgrad.manchester.ac.uk}}

\institute{Jodrell Bank Centre for Astrophysics, School of Physics and Astronomy,	The University of Manchester, Manchester, M13 9PL, U.K.
	\and School of Physics, University of Exeter, Exeter EX4 4QL, U.K.
  \and Department of Physics, Bucknell University, 
  Lewisburg, PA 17837, U.S.A.
  \and Astrophysics Group, Cavendish Laboratory, J J Thomson Avenue, Cambridge, CB3 0HE, U.K.
  \and Kavli Institute for Cosmology, c/o Institute of Astronomy, University of Cambridge, Madingley Road, Cambridge, CB3 0HA, U.K.}
  
\date{Accepted for publication in A\&A on 17 May 2010}

\abstract 
{The Serpens North Cluster is a nearby low mass star forming region which is
  part of the Gould Belt. It contains a range of young stars thought
  to correspond to two different bursts of star formation and provides the
  opportunity to study different stages of cluster formation.}
{This work aims to study the molecular gas in the Serpens North Cluster to
  probe the origin of the most recent burst of star formation in Serpens.  }
{Transitions of the C$^{17}$O and C$^{18}$O observed with the IRAM 30m
  telescope and JCMT are used to study the mass and velocity structure of the
  region while the physical properties of the gas are derived using LTE and
  non-LTE analyses of the three lowest transitions of C$^{18}$O.}
{The molecular emission traces the two centres of star formation which are
  seen in submillimetre dust continuum emission.  In the $\sim$40~M$_\odot$ NW
  sub-cluster the gas and dust emission trace the same structures
  although 
  there is evidence of some depletion of the gas phase C$^{18}$O. The gas has
  a very uniform temperature ($\sim$10~K) and velocity ($\sim$8.5~kms$^{-1}$)
  throughout the region.  This is in marked contrast to the SE sub-cluster. In
  this region the dust and the gas trace different features, with the
  temperature peaking between the submillimetre continuum sources, reaching up
  to $\sim$14~K. The gas in this region has double peaked line profiles which
  reveal the presence of a second cloud in the line of sight. The
  submillimetre dust continuum sources predominantly appear located in the
  interface region between the two clouds.}
{ Even though they are at a similar stage of evolution, the two Serpens
  sub-clusters have very different characteristics. We propose that these
  differences are linked to the initial trigger of the collapse in the regions
  and suggest that a cloud-cloud collision could explain the observed
  properties.}

\maketitle


\section{Introduction}
\label{intro}


\label{serpens}

Despite the importance of understanding the processes driving the formation of
stars in the Galaxy, little is known about the role played by molecular cloud
kinematics on triggering or suppressing star formation.  Since most stars
form in clusters \citep{2003ARA&A..41...57L}, the kinematics of young stellar
clusters in which the initial conditions of clustered star formation are still
imprinted in the gas and dust emission properties can provide important
insights into the dominant mode of star formation \citep[e.g.][]{2006A&A...445..979P}.

\begin{table*}[!ht]
	\footnotesize
	\centering
	\caption{\small Submillimetre sources in Serpens Main Cluster}
		\begin{tabular}{c c c c c}
		\hline 
		\hline
		Source name & RA (J2000) & Dec (J2000) & offset RA ($''$) & offset Dec ($''$)\\ 
		\hline
		SMM 1 &  18:29:49.87 & 1:15:16.0 & 0.3 & 2.6 \\  
		SMM 2 &  18:30:00.25 & 1:12:51.7 & 0.8 & 5.7 \\
		SMM 3 &  18:29:59.26 & 1:13:56.3 & 0.3 & 2.0 \\
		SMM 4 &  18:29:56.77 & 1:13:08.0 & 2.7 & 2.1 \\
		SMM 5 &  18:29:51.35 & 1:16:34.9 & 3.3 & 0.9 \\ 
		SMM 6 &  18:29:57.99 & 1:13:59.2 & 4.7 & 3.0 \\
		SMM 8 &  18:30:01.88 & 1:15:08.4 & 0.5 & 0.9 \\
		SMM 9 &  18:29:48.34 & 1:16:42.0 & 3.3 & 0.5 \\
		SMM 10 & 18:29:52.04 & 1:15:44.4 & 1.5 & 3.4 \\
		SMM 11 & 18:30:00.41 & 1:11:41.6 & 1.4 & 0.8 \\
		\hline
		\end{tabular}
	\label{tab:SMMsources}
\end{table*}

One such young and nearby cluster is in the Serpens Molecular Cloud
(MC). Located at $\sim$260~pc \citep{1996BaltA...5..125S}, the optical
extinction map of the cloud covers more than 10~deg$^{2}$
\citep{1999A&A...345..965C}. However, the majority of the star formation is
occuring in three clusters covering approximately 1.5~deg$^{2}$
\citep{2007ApJ...666..982E}.  The most active region is the Serpens Main
Cluster (hereafter Serpens) which has a surface density of YSOs of
222~pc$^{-2}$, compared to 10.1~pc$^{-2}$ in the rest of the Serpens cloud
\citep{2007ApJ...663.1149H}. In this main cluster, the average gas density is
around 10$^{4}$~cm$^{-3}$ \citep{2007ApJ...666..982E} with H$_{2}$ column
densities greater than 10$^{22}$~cm$^{-2}$ in the cores.  The high density of
protostars in this main cluster seems to indicate an early stage of evolution
where the cluster gas may still be infalling into the cores
(\citealt{1999ApJ...518L..37W}; \citeyear{2000ApJ...537..891W};\citealt{1996ApJ...456..686H}).
The star formation rate in this main cluster is
56~M$_{\odot}$Myr$^{-1}$pc$^{-2}$, $\sim20$ times higher than in the rest of
the cloud \citep[][]{2007ApJ...663.1149H}.

\begin{figure}[!h]
	\centering
	\includegraphics[width=0.45\textwidth]{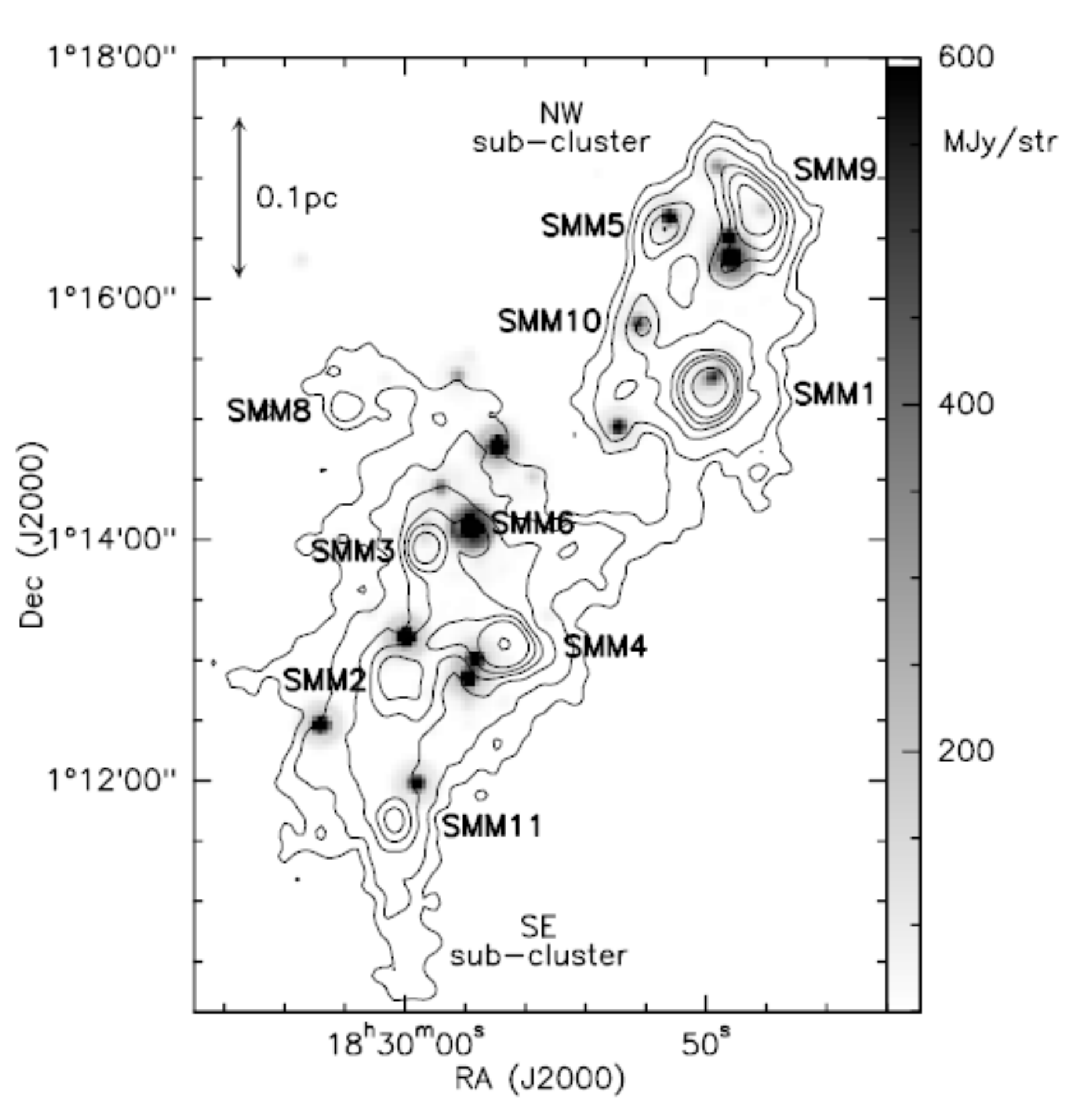}
	\caption{\small{Map of the SCUBA 850~$\mu$m continuum emission in
            contours showing the position of the submillimetre sources
            (labeled). Contours at 0.4, 0.6, 1, 1.4, 1.8, 2.4 and
            5~Jy~beam$^{-1}$. In grey scale are the Spitzer MIPS 24~$\mu$m
            sources \citep{2007ApJ...663.1139H}. All the sources seen on this
          figure are classified as being young protostars, mostly Class O and
          Class I sources, with a small number of flat spectrum sources.}}
	\label{fig:srp}
\end{figure}

Amongst the youngest YSOs found in Serpens there are ten Class 0 and I
protostars which are detected in 850~$\mu$m dust continuum emission
(e.g. \citealt{1996ApJ...460L..45H}, \citealt{1999MNRAS.309..141D}), hereafter
referred to as submillimetre sources (shown on Fig.~\ref{fig:srp},
Table~\ref{tab:SMMsources} and discussed on Sec.~\ref{scuba.data}). These are
distributed within $\sim$~0.2~pc$^2$ and divided between two sub-clusters, one
to the northwest (NW) and one to the southeast (SE). These submillimetre
sources power a number of outflows, which have been studied using several
different approaches
\citep[e.g.][]{1992PASJ...44..155E,1999MNRAS.309..141D,1999AJ....118.1338H,2000ApJ...530L.115D,2010MNRAS...Graves}. {Figure~\ref{fig:srp} also shows the Spitzer
  MIPS 24~$\mu$m emission tracing the young protostars classified as Class I
  or 0. The oldest objects in the area shown on the image are a few flat
  spectrum sources \citep{2007ApJ...663.1149H,2004A&A...421..623K}.}  The
presence of Class II and Class III objects {(not shown in
  Figure~\ref{fig:srp})} dispersed over a larger area suggests that the region
has undergone two episodes of star formation. {The first, responsible for
  these dispersed pre-main sequence stars (the Class II and III sources),
  occurred about 2~Myr before the most recent burst which formed the
  submillimetre and 24$\mu$m protostars (Class 0, I and flat spectrum
  sources), $~10^5$~yr ago
  \citep[][]{2007ApJ...663.1149H,2004A&A...421..623K}.}

This paper focuses on the dynamical and physical properties of the gas in
Serpens using CO isotopologues observed with the IRAM 30m telescope and with
JCMT, to probe the current properties of the region as well as investigate any
link back to the initial conditions under which the most recent burst of star
formation in Serpens took place. Section 2 presents the observations,
as well as the data reduction and analysis methods and techniques used in this
study. 
In Section 3 the structure of the gas is discussed while Section 4 discusses
its dynamics. In Section 5 its physical properties are analysed. These results
are drawn together and a scenario for the origin of the star formation in
Serpens described in Section 6.

\section{Data and Analysis Techniques}
\label{data}

\subsection{IRAM Observations}
\label{iramobserv}

The Serpens region was observed in the J=1$\rightarrow$0 and J=2$\rightarrow$1
transitions of C$^{18}$O and the J=1$\rightarrow$0 transition of C$^{17}$O with
the IRAM 30m telescope, using the facility receivers, in May 2001.  The
observations consisted of on-the-fly maps of the region, centered at RA =
$18^{h}$$29^{m}$$57.91^{s}$ and Dec = $1^{\circ}$$12'$$25.2''$ over an area of
approximately 10.5~arcmin$^{2}$, $\sim$~$3'$ in Right Ascension and $3.5'$ in
Declination. 

The C$^{17}$O J=1$\rightarrow$0 data, observed at 112.359~GHz, have a spatial
resolution of $22''$, a velocity resolution of $\sim$~0.052~kms$^{-1}$ and a
noise level of $\sim$~0.45~K (in T$_{\mathrm{A}}^{*}$) in the raw map -- low
enough to allow the detection and identification of the hyperfine components
of the J=1$\rightarrow$0 transition of C$^{17}$O. C$^{18}$O was observed with
spectral resolution of $\sim$~0.053~kms$^{-1}$ at 109.782~GHz and 219.816~GHz
and with spatial resolution of $22''$ and $11''$ for the J=1$\rightarrow$0 and
J=2$\rightarrow$1 transition, respectively.  Both emission lines are detected
with a good signal to noise, both with a one sigma noise level of
$\sim$~0.45~K in T$_{\mathrm{A}}^{*}$.

The beam and forward efficiencies of the IRAM 30m telescope
(B$_{\mathrm{eff}}$ and F$_{\mathrm{eff}}$ respectively) are given on the
telescope website. From these we estimate for both C$^{17}$O and C$^{18}$O, a
F$_{\mathrm{eff}}$=0.95 and B$_{\mathrm{eff}}$=0.72, for the J=1$\rightarrow$0
transition, and a F$_{\mathrm{eff}}$=0.91 and B$_{\mathrm{eff}}$=0.54, for the J=2$\rightarrow$1
transition.

The main data reduction was performed using GILDAS software (CLASS90 and
GREG). This included the baseline corrections, hyperfine/gaussian fitting of
the data, and construction of the datacubes. Given the good quality of the data,
the baselines were well fitted by a simple first degree polynomial function.

\subsubsection{C$^{17}$O}
\label{c17odata}

\begin{figure*}[!t]

	\centering
	\includegraphics[width=\textwidth]{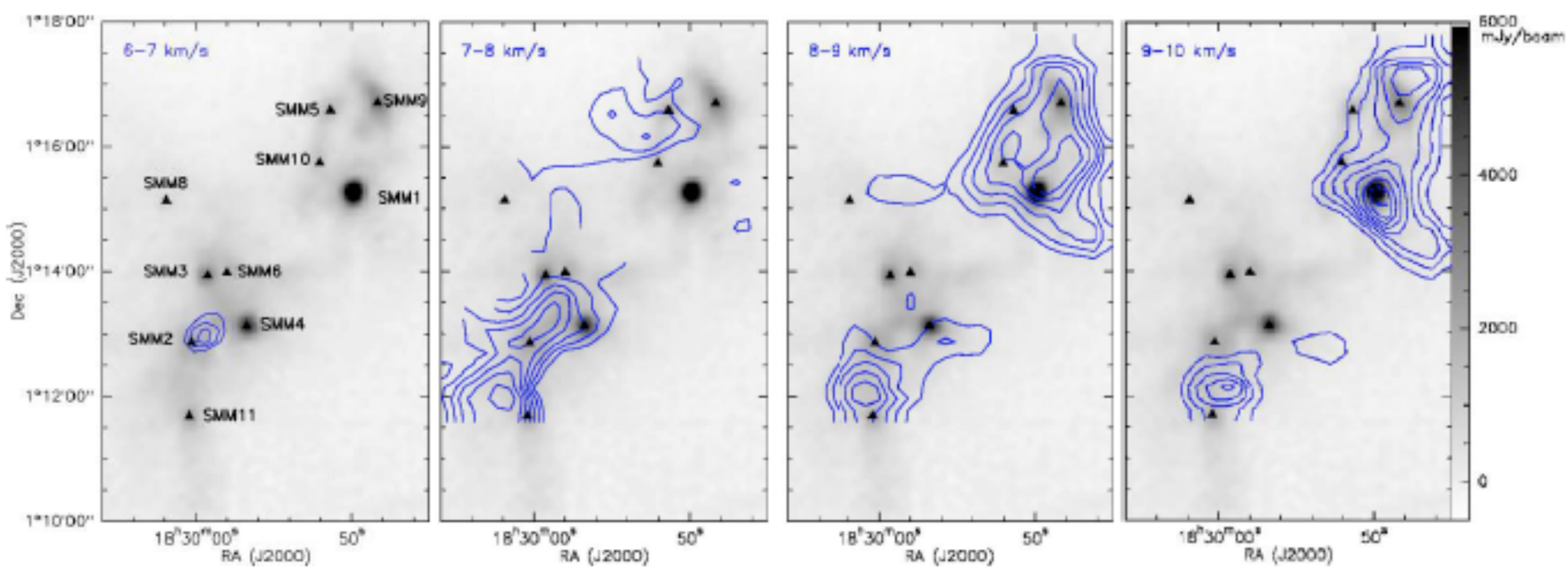}
	\caption{\small{Contour maps of C$^{17}$O J=1$\rightarrow$0 modelled
            emission from the hyperfine fit, overplotted on SCUBA 850$\mu$m
            emission in gray scale. The submillimetre sources are identified
            on the left figure with labels and triangles. These symbols will
            be used from this point forward. Each figure represents a mean
            intensity map over 1~kms$^{-1}$ intervals, from 6 to
            10~kms$^{-1}$. Contours range from 0.4~K\,kms$^{-1}$ increasing by
            steps of 0.1~K\,kms$^{-1}$ (in T$_{\mathrm{A}}^{*}$ scale). The
            figure shows the SE sub-cluster with three main peaks: one above
            SMM11, one between SMM2 and SMM6, and finally, one west of
            SMM4. While the NW sub-cluster also contains three major peaks:
            one on SMM1, one close to SMM9, and another one south-west of
            SMM1.}}
  \label{fig:channels}	          
\end{figure*}

The C$^{17}$O J=1$\rightarrow$0 line comprises three, partially blended,
hyperfine features.  By fitting the hyperfine structure (HFS) of the spectrum,
the line width, velocity and optical depth ($\tau$) can be extracted. The line
shape in the presence of hyperfine structure can be described by
\begin{equation}
T(v) = T_{S}(1-e^{-\tau(v)}),
\label{eq:hfs}
\end{equation}
where 
\begin{equation}
\tau(v) = \tau_{0} \sum^{3}_{i=1} r_{i}
\exp \biggl({-\frac{(v-v_{0,i})^{2}}{2\sigma^{2}}}\biggr), 
\end{equation}
$T(v)$ is the line brightness temperature, $T_{S}$ is the source temperature
and $\tau$ is the optical depth. The optical depth is the sum over the three
hyperfine components of the transition with $r_{i}$ and $v_{o,i}$ - the
relative weight and the central velocity for each hyperfine component - being
$v$ the velocity, $\sigma$ the velocity dispersion and $\tau_{0}$ the total
optical depth common to the three components
\citep[][]{1993ApJ...418..273F}. The spacing and weight of the hyperfine
components were adopted from \citet{1998ApJ...495..871L}.  Further details
about the hyperfine structure fitting procedure in the GILDAS software can be
found on the IRAM
website\footnote{\textit{http://www.iram.fr/IRAMFR/GILDAS/doc/html/class-html/node8.html}}.

To fit this hyperfine structure, the individual spectrum at each pixel in the
image was extracted from the datacube and was fitted using the procedure
described above. A model Gaussian spectrum for each pixel was then
reconstructed using the derived values (the peak intensity, line width and
central velocity).  Only pixels where both the line width and line peak
intensity were determined with a signal to noise ratio of 5 or greater were
considered.

The initial fitting showed that {within the uncertainties, all} the
emission was {consistent with being} optically thin.  Therefore, to
reduce the uncertainties on the fitted quantities, the HFS fitting was redone
fixing the $\tau$ at 0.1 for the whole map, {consistent with optically
  thin emission}. In the final C$^{17}$O J=1$\rightarrow$0 modelled datacube
(Fig.~\ref{fig:channels}) we were able to identify clear peaks at different
velocities and positions in the region. A detailed study of these peaks is presented in Section
\ref{c17o.res}.


\subsubsection{C$^{18}$O}
\label{c18odata}

\begin{figure}[!h]
	\centering
	\includegraphics[width=0.5\textwidth]{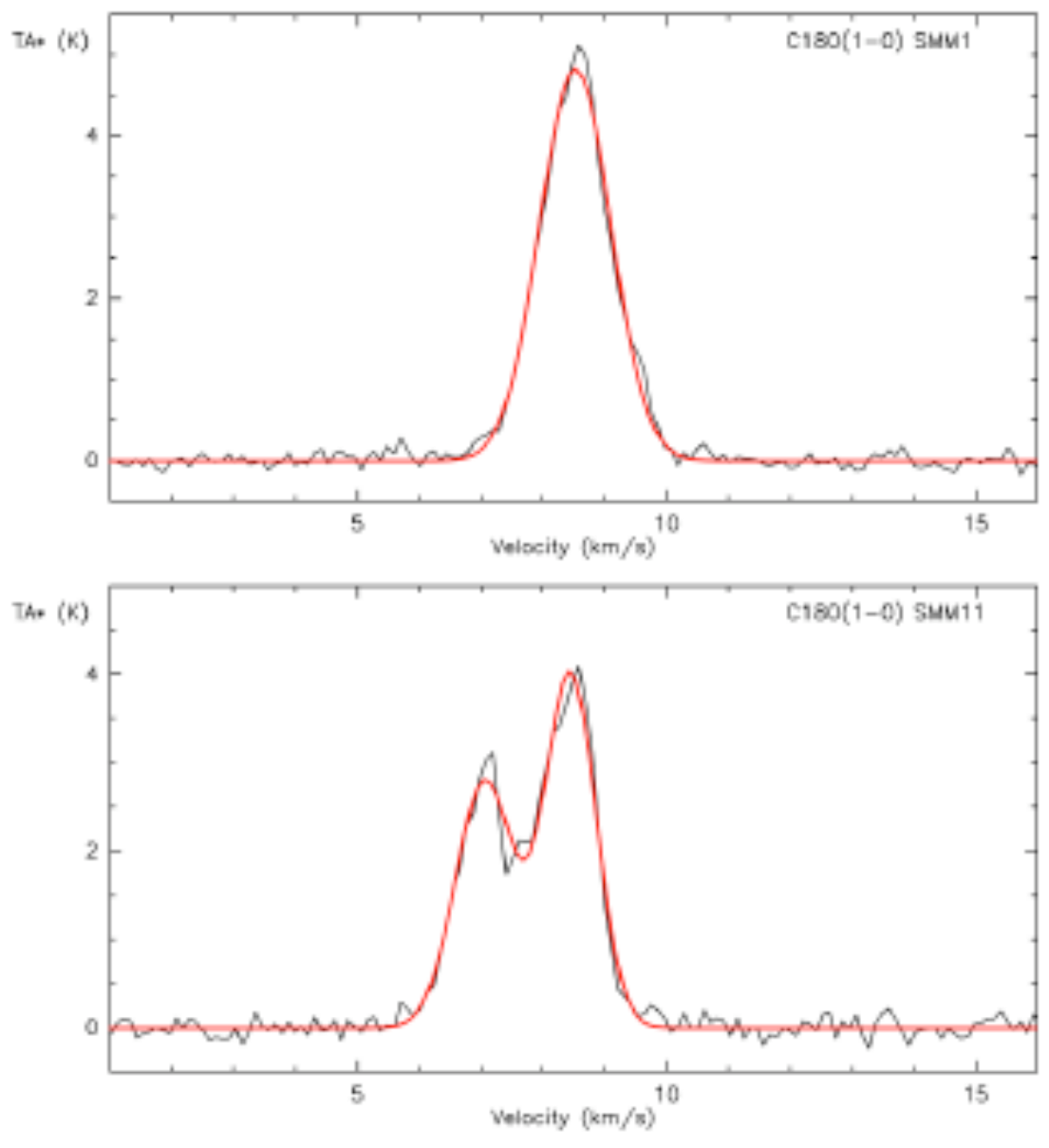}	
	\caption{\small{Observed C$^{18}$O J=1$\rightarrow$0 spectra (black
            solid line), smoothed in velocity to 0.1~kms$^{-1}$ width
            channels, at two different positions: on SMM1 (top) and on SMM11
            (bottom); with the respective Gaussian fit (red line). {These are
              examples of a single peaked spectrum as seen in the NW
              sub-cluster and a double peaked profile, as seen in the SE
              sub-cluster.}}}
	\label{fig:c18ospec}
\end{figure}

The Serpens C$^{18}$O emission is not affected by the outflows in the region
(Section \ref{ss:outflows}) and the C$^{18}$O lines have no hyperfine
structure. Therefore, on some regions of the mapped cloud such as the NW
sub-cluster (Fig.~\ref{fig:srp}), lines are well represented by a single
Gaussian (Fig.~\ref{fig:c18ospec} left panel). However, in the SE sub-cluster,
the line profile has two clear peaks (Fig.~\ref{fig:c18ospec} lower
panel). The low optical depth of the C$^{18}$O emission
(Section~\ref{c17odata} and \ref{scatter}) together with double peaked lines
in this region in other optically thin tracers such as N$_{2}$H$^{+}$
J=1$\rightarrow$0 \citep[][]{2002A&A...392.1053O} suggests that these two
peaks trace two clouds along the line of sight towards this sub-cluster.  In
the weaker C$^{17}$O emission with its blended hyperfine structure, these two
velocity components are too difficult to separate.
However, since it is optically thin, the C$^{17}$O emission is still a
reliable tracer of the total column density.

\subsection{JCMT data}
\label{harp}

Our excitation analysis (Section~\ref{c18o.res}) makes use of JCMT HARP data
from the Gould Belt Survey (GBS) at JCMT \citep{2010MNRAS...Graves,2007PASP..119..855W}.  
The data used is C$^{18}$O J=3$\rightarrow$2,
at 329.330~GHz, with 0.055~kms$^{-1}$ spectral resolution, and $14''$ spatial
resolution. The telescope main beam efficiency at this frequency is
$\eta_{\mathrm{mb}} = 0.66$ \citep{2009MNRAS.tmp.1566C}, and the rms level
achieved is of the order of 0.2~K (T$_{\mathrm{A}}$).  A full description of
these data is given in the GBS Serpens First Look paper (\citealt{2010MNRAS...Graves}, 
hereafter referred to as SFLPaper).

\label{scuba}

The submillimetre continuum data at 850~$\mu$m was observed with SCUBA at the
JCMT, with a beam size of $14''$. The initial reduction, analysis
and discussion of these data was presented by \citet{1999MNRAS.309..141D},
where they estimate the overall dust properties and  characteristics
of the cloud (see Section \ref{serpens}).  We have used the pipeline reduced
SCUBA data from the Canadian Astronomy Data Centre (CADC) 
archives\footnote{http://www.cadc.hia.nrc.gc.ca/jcmt/} 
to investigate the structure of the dust continuum emission and for 
comparison with the IRAM 30m C$^{17}$O and C$^{18}$O data.


\label{scuba.data}

An initial inspection of the SCUBA data indicated good agreement in the source
positions for those sources where \citeauthor{1999MNRAS.309..141D} determined
positions from this same SCUBA data (SMM8 and SMM11) and as well as for SMM3.
{However, in agreement with interferometric continuum observations
  \citep{1999ApJ...513..350H}, the positions of some of the remaining SMM
  sources needed to be revised compared to those listed in
  \citet{1999MNRAS.309..141D} with absolute offsets from the published
  positions greater than $5''$ for SMM2 and SMM6}.  Table
\ref{tab:SMMsources} presents redetermined positions for all sources,
extracted from the 850~$\mu$m map of Serpens, which now agree within 1$''$ of
the positions in the SCUBA cores catalogue published by
\citet{2008ApJS..175..277D}.
We estimate that these positions are accurate within the 2$''$ SCUBA pointing
errors \citep[][]{1999MNRAS.309..141D}.
The offsets in RA and Dec between the revised positions and those previously
published \citep[listed in][]{1999MNRAS.309..141D} are also shown on Table
\ref{tab:SMMsources}.


\section{Gas structure of the cloud} 
	\label{general.structure}
\label{res}

To determine the structure of the molecular gas we have carried out a clumping
analysis in 2D and 3D. Using the velocity information from the gas emission it
is possible to identify the individual clumps within the cloud. These are
compared to the structure visible in the dust continuum.  We use this analysis
to quantify the sizes and masses of molecular gas associated with protostars,
and carry out a virial analysis to determine the clump stability.

\subsection{C$^{17}$O  2D-Clumps}
\label{c17o.res}

Initially we manually extracted the small scale molecular structures for
comparison with the dust seen in the SCUBA map. This was done based on a
visual inspection of both channel and integrated intensity maps
(e.g. Fig.~2). The C$^{17}$O J=1$\rightarrow$0 channel maps show a significant
number of emission features which are not directly associated with the SCUBA
cores. For this reason we call these molecular structures ``clumps" although
this term is often used to describe parsec-scale structures
\citep{1993prpl.conf..125B}.

\begin{table*}[!ht]
	\footnotesize
	\caption{\small Properties of the 2D-clumps}
		\begin{tabular}{c c c c c c c c c c}
		\hline 
		\hline
		2D-Clump & RA$_{\mathrm{peak}}$  & Dec$_{\mathrm{peak}}$ & V$_{\mathrm{peak}}$ & Area  & FWHM  & M$_{\mathrm{cf}}$ & M$_{\mathrm{virial}}$ & Ratio & $I_{\mathrm{low}}$ \ \ \ \ $I_{\mathrm{peak}}$\\
		ID & (J2000) & (J2000) & (kms$^{-1}$) & (arcmin$^{2}$) & (kms$^{-1}$) & (M$_{\odot}$) & (M$_{\odot}$) & (M$_{\mathrm{virial}}$/M$_{\mathrm{cf}}$) & (K\,kms$^{-1}$)\\
		\hline
		A	& 18:29:49.89  & 01:15:15 & 8.61 & 2.30 & {1.1} & {9.0}	 & {9.3}  & {1.0} & 1.15 \ \ \ \ {2.42}\\
		B	& 18:29:48.43  & 01:17:04 & 8.55 & 1.83 & {1.1} & {5.4}  & {8.3}  & {1.5} & 0.90 \ \ \ \ {1.90}\\
		C	& 18:29:46.97	 & 01:14:31 & 8.65 & 1.33 & {1.4} & {4.7}	 & {12.9}	& {2.7} & 1.10 \ \ \ \ {2.17}\\
		D &	18:29:55.43  & 01:16:25 & 7.73 & 0.98 & {1.3} & {1.9}	 & {11.8} & {6.2} & 0.50 \ \ \ \ {0.74}\\
		E	& 18:29:59.40	 & 01:13:04 & 7.43 & 1.00 & {2.2} & {2.0}	 & {25.2}	& {12.6}& 0.60 \ \ \ \ {1.99}\\
		F	& 18:30:00.87	 & 01:11:58 & 8.29 & 1.63 & {1.8} & {7.1}	 & {23.4}	& {3.3} & 1.35 \ \ \ \ {2.89}\\
		G	& 18:29:55.75	 & 01:12:53 & 8.36 & 0.70 & {1.4} & {1.6}	 & {8.5}	& {5.3} & 0.90 \ \ \ \ {1.33}\\
		NW& 18:29:49.89  & 01:15:15 & 8.61 & 9.06 & {1.2} & {31.3} & {33.2}	& {1.1} & 1.00 \ \ \ \ {2.42}\\
		SE& 18:30:00.87  & 01:11:58 & 8.29 & 7.49 & {1.9} & {26.0} & {70.3}	& {2.7} & 1.00 \ \ \ \ {2.89}\\
		\hline
		\end{tabular}
	\label{tab:2D-clumps}
\end{table*}

\begin{figure}[!t]
	\centering
	\includegraphics[width=0.48\textwidth]{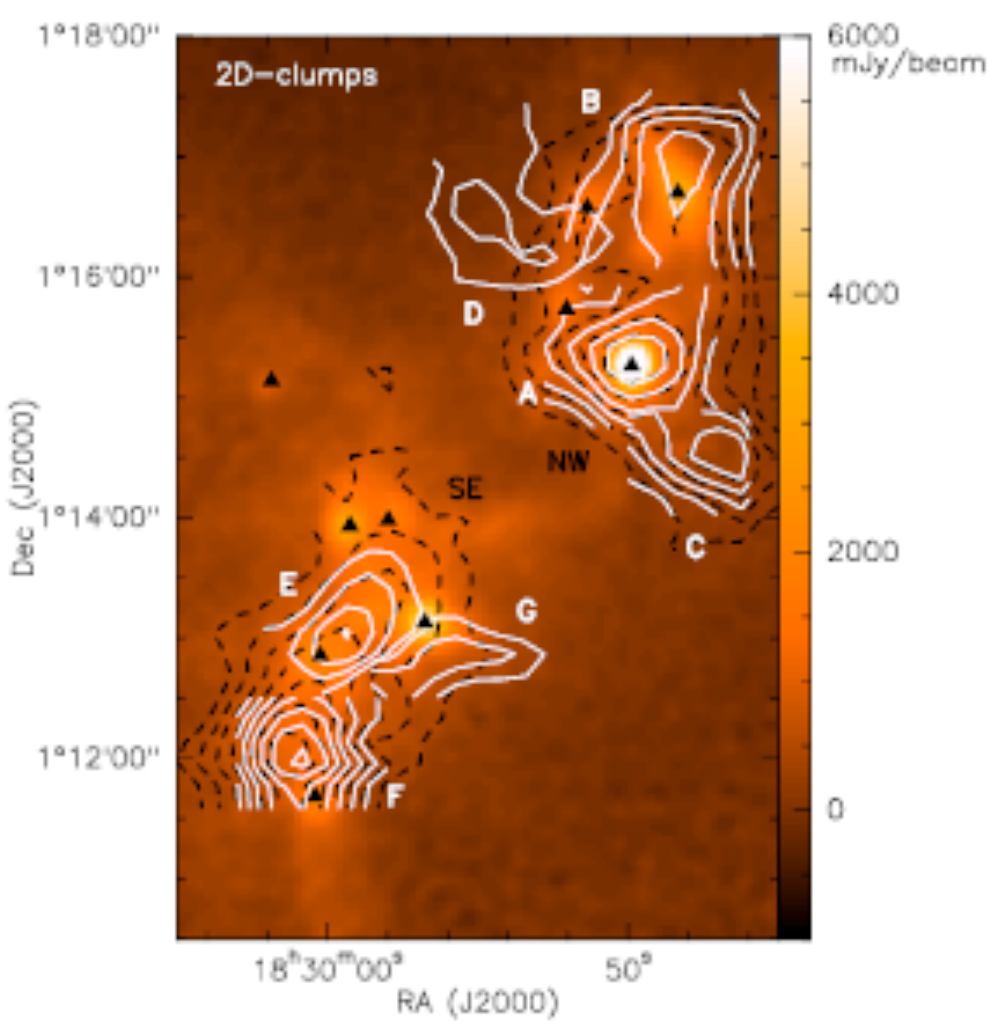}
	\caption{\small SCUBA map of the 850~$\mu$m continuum emission (colour
          scale) overplotted with the positions of the submillimetre sources
          (triangles), the C$^{17}$O J=1-0 2D-clumps (solid white contours and
          letters) and the NW and SE sub-clusters (black dashed contours). The
          solid white contours are the intensity integrated over the
          correspondent velocity range for each clump. These are stepped by
          0.2~K\,kms$^{-1}$, except for weaker 2D-clumps G and D, stepped by
          0.1~K\,kms$^{-1}$ (in T$_{A}^{*}$). The lower level of each clump is
          the same as the specified on Table \ref{tab:2D-clumps}. The NW and
          SE sub-clusters contours are integrated intensity over the entire
          velocity range. Contours are stepped by 0.25~K\,kms$^{-1}$, with the
          lower contour at 1.25~K\,kms$^{-1}$ (in T$_{\mathrm{A}}^{*}$).}
	\label{fig:2D-clumps}
\end{figure}

The properties of each identified clump (Fig.~\ref{fig:2D-clumps}) was
subsequently extracted using the IDL 2D version of the source extraction
\textsc{clumpfind} algorithm code by \citet{1994ApJ...428..693W} on maps
integrated over the velocity range in which each clump appeared.
 
With the size and the integrated intensity {corrected for telescope efficiency}, we estimated the column density and
mass of each clump (M$_{\mathrm{cf}}$) assuming a temperature of 10~K, a mean molecular weight of 2.33 and a
C$^{17}$O fractional abundance with respect to H$_{2}$ of
4.7~$\times$~10$^{-8}$ \citep{1982ApJ...262..590F,2002A&A...389..908J}.
We also calculated the clumps virial masses using Eq.~\ref{eq:virialmass},
where $M_{\mathrm{vir}}$ is the virial mass, $\sigma_{\mathrm{obs}}$ is the
observed velocity dispersion, $G$ is the gravitational constant and $\alpha$
is a coefficient function of the adopted density profile: $\alpha$ is 3/5 for
a uniform density, 2/3 for a profile as $\rho \propto r^{-1}$, 3/4 when $\rho
\propto r^{-1.5}$, and 1 when $\rho \propto r^{-2}$.

\begin{equation}
M_{\mathrm{vir}} = \frac{3 R \sigma_{\mathrm{obs}}^{2}}{\alpha G}
\label{eq:virialmass}
\end{equation}

The listed virial masses of the clumps adopt a
density profile of $\rho \propto r^{-2}$.
{The velocity FWHM of each clump was estimated by averaging 
all the spectra assigned to that clump} and has an estimated 
uncertainty of $\sim$0.1~kms$^{-1}$.  The clumps identified by this
method will be referred to as the 2D-clumps hereafter.

The virial mass (M$_{\mathrm{virial}}$) and the gas mass (M$_{\mathrm{cf}}$) 
were also calculated for the two sub-clusters, NW and SE. The method was the same as
for the clumps except the density profile for the sub-cluster gas was assumed
to be $\rho \propto r^{-1.5}$, which is expected to be more appropriate for
these larger size regions. If the same $\rho \propto r^{-2}$ as for the clumps
had been adopted, the derived sub-cluster mass would be a factor of 25\%
smaller.

The clumps are shown on Fig.\ref{fig:2D-clumps}, and the physical parameters
summarized on Table \ref{tab:2D-clumps}, where: RA$_{\mathrm{peak}}$ and
Dec$_{\mathrm{peak}}$ are the position where the emission peaks within each
clump; V$_{\mathrm{peak}}$ is the velocity at the peak position, with an
uncertainty of 0.05~kms$^{-1}$; area is the surface in the map occupied by
each clump; M$_{\mathrm{cf}}$ is the mass of the clump calculated from
\textsc{clumpfind} outputs; M$_{\mathrm{virial}}$ is the virial mass; Ratio
M$_{\mathrm{virial}}$/M$_{\mathrm{cf}}$ is a measurement of how bound each
clump is - a gravitationally bound structure should have a ratio around unity,
but given the uncertainties of these calculations, we consider a structure to
be unbound if the ratio is above 2; I$_{\mathrm{low}}$ represents the lower
contouring level assumed when running the algorithm for each different clump
(increasing with steps of 0.10~K\,kms$^{-1}$); {and, finally,
  I$_{\mathrm{peak}}$ shows the integrated intensity in T$_{\mathrm{A}}^{*}$as
  measured at the peak position}.

{Observational sources of uncertainty include the distance to Serpens and the
  line width. Uncertainties on the line width in particular might be a special
  issue in the SE region where the two velocity components observed in
  C$^{18}$O may become important in broadening the C$^{17}$O line.  Systematic
  uncertainties in M$_{\mathrm{cf}}$ include uncertainty in the adopted gas
  temperature and fractional abundance of C$^{17}$O. Finally, the systematic
  uncertainties on the M$_{\mathrm{virial}}$ include source geometry effects
  and the neglection of additional terms in the virial equation (due to
  external pressure, magnetic pressure, etc.).  Amongst all the possible
  sources of uncertainty, the greatest is likely to be the factional abundance
  of C$^{17}$O, given that our non-LTE study of C$^{18}$O at 8 positions
  (Sec.~\ref{scatter}) show a mean depletion factor of 2.5
  (App.~\ref{intensities}).  Given the observational and possible systematic
  uncertainties on the calculations,
  the virial ratio is perhaps best seen as a useful tool to compare the different structures
  within a cloud rather than absolute measure of the gravitational equilibrium
  of any given clump. }

The NW and SE sub-clusters are extended regions, and therefore, the peak
positions and velocities correspond to one of the smaller identified clumps
lying within the sub-cluster. The NW sub-cluster peaks at the position of
clump A (\& SMM1) and the SE sub-cluster peaks at the position of clump F
(north of SMM11). Similarly, the velocities quoted for the peak for the
sub-clusters are not the mean velocity of the sub-clusters, but the velocity
at the peak of the strongest clump.

Note that even though both sub-clusters, SE and NW, have similar masses
(M$_{\mathrm{cf}}$), they each have a different equilibrium status, with a
factor of 3 difference between their respective virial ratio. Interestingly,
{even when considering some depletion (Appendix~~\ref{intensities})}, the SE
region is {likely} super-virial whereas the NW, due to its smaller line width,
is {marginally} sub-virial. About 67\% of the mass in the NW region and 40\%
of the mass in the SE region is associated with the clumps. {Four of the
  clumps (A, B, C and F) are individually within a factor of three of being in
  virial equilibrium. Accounting for some depletion of C$^{18}$O
  (Appendix~~\ref{intensities}), M$_{\mathrm{cf}}$ could increase up to a
  factor of 2.5, which would make all these four clumps relatively bound
  structures.  Clump D is a factor of $\sim6$ super-virial, and is likely to
  be less affected by depletion as the dust densities are lower, likely
  identifying this clump as part of a more diffuse region which is less bound
  than the NW sub-cluster.  Finally, even accounting for possible depletion,
  clumps, E and G, with mass ratios of $\sim13$ and $\sim5$, are likely
  unbound structures.}  They may either represent shocked regions where the
line width is intrinsically high (from 1.6 to 2.2~kms$^{-1}$) or regions
where, as mentioned in section \ref{c18odata}, two blended velocity components
contribute to the emission along the same line of sight. The HFS fitting of a
single component in this case would result in a broadening of the line width
due to blending of the two components. However, in Sec.~\ref{sec:decomp} we
see that even after separating the components, the line width of at least one
of them is still broader than seen anywhere in the NW, pointing to a genuine
broad line emission. {In summary, the SE sub-cluster is much more
  dynamic  than the NW, with a kinematic support a few times
  higher, both when comparing individual clumps and the overall sub-clusters.}


\subsection{C$^{17}$O 3D-Clumps}
\label{c17o3d.res}

Although the 2D clumpfinding is valuable for comparison with the dust
continuum, it is limited in its ability to represent the true structure of the
cloud. The 3D \textsc{clumpfind} automatically studies the datacube in all
three dimensions of space-space-velocity. In particular, 3D \textsc{clumpfind}
should provide a better understanding of the cloud's structure where clumps
may overlap along a line of sight but have different velocities, or where the
emission is narrow in velocity making it weak in integrated intensity maps.
Therefore we have complemented the 2D study of the structure of the C$^{17}$O
using the 3D version of the \textsc{clumpfind} 
within the Starlink package.

\begin{table*}[!t]
\begin{center}
	\footnotesize
	\caption{\small Properties of the 3D-clumps}
		\begin{tabular}{c c c c c c c c c c}
		\hline 
		\hline
		3D-clump  & RA$_{\mathrm{peak}}$  & Dec$_{\mathrm{peak}}$ & V$_{\mathrm{peak}}$ & Area  & FWHM & M$_{\mathrm{cf}}$ & M$_{\mathrm{virial}}$ & Ratio & T$_{\mathrm{peak}}$\\
		ID  & (J2000) & (J2000) & (kms$^{-1}$) & (arcmin$^{2}$) & (kms$^{-1}$) & (M$_{\odot}$) & (M$_{\odot}$) & (M$_{\mathrm{virial}}$/M$_{\mathrm{cf}}$) & (K)\\
  	\hline
	 1	& 18:29:51.0 & 1:15:04  & 8.64 & 3.93 & 1.0 & {9.0}	 & 10.5 & {1.2} & {2.0}\\
	 2	& 18:29:49.2 & 1:16:09  & 8.56 & 3.46 & 1.4 & {7.3}	 & 20.0 & {2.7} & {1.6}\\ 
	 3	& 18:30:01.6 & 1:11:47  & 8.67 & 2.30 & 2.0 & {6.3}	 & 31.2 & {4.9} & {1.3}\\ 
	 4	& 18:29:49.2 & 1:09:57  & 8.20 & 1.43 & 0.3 & {0.6}	 & 0.4  & {0.7} & {1.1}\\ 
	 5  & 18:29:47.0 & 1:13:58  & 8.80 & 1.53 & 1.2 & {2.5}	 & 9.1  & {3.6} & {1.0}\\ 
	 6	& 18:29:55.0 & 1:12:41  & 8.51 & 1.13 & 1.1 & {1.3}	 & 6.4  & {4.9} & {0.9}\\ 
	 7	& 18:30:10.3 & 1:13:25  & 8.09 & 1.00 & 0.4 & {0.6}	 & 1.0  & {1.7} & {1.1}\\ 
	 8	& 18:29:44.8 & 1:17:04  & 8.80 & 1.77 & 0.9 & {1.2}  & 5.3  & {4.4} & {1.1}\\ 
	 9  & 18:29:47.0 & 1:09:24  & 8.41 & 0.73 & 0.4 & {0.3}	 & 0.6  & {2.0} & {1.0}\\
	 10 & 18:30:00.1 & 1:12:52  & 7.68 & 1.13 & 1.5 & {1.6}	 & 13.7 & {8.6} & {0.8}\\
	 11 & 18:29:58.7 & 1:15:04  & 8.23 & 0.73 & 0.9 & {0.8}	 & 4.0  & {5.0} & {0.8}\\ 
   12 & 18:30:06.0 & 1:11:58  & 7.36 & 1.07 & 1.2 & {0.8}	 & 8.4	& {10.5} & {0.7}\\
	 13 & 18:29:57.9 & 1:15:48  & 7.88 & 1.67 & 0.6 & {0.7}	 & 2.6	& {3.7} & {0.7}\\ 
	 14 & 18:29:57.2 & 1:11:47  & 8.67 & 0.90 & 0.8 & {0.4}	 & 3.0  & {7.5} & {0.7}\\ 
	 15 & 18:29:55.7 & 1:16:20  & 7.73 & 1.37 & 0.8 & {0.6}	 & 3.9  & {6.5} & {0.8}\\ 
	 16 & 18:29:47.0 & 1:11:48  & 8.80 & 0.93 & 0.3 & {0.2}	 & 0.3	& {1.5} & {0.7}\\ 
	 \hline
		\end{tabular}
		\end{center}
	\label{tab:3D-clumps}
\end{table*}

{Similarly to \citet{2009ApJ...699L.134P}, we also found that the results
  on} the 3D \textsc{clumpfind} analysis to be very sensitive to the
parameters used, especially in characterising the weaker emitting regions.
Stronger clumps were unequivocally detected with a wide range of parameters,
but changing the step size and/or the number of pixels per clump allowed to be
adjacent to a bad pixel would result in the merging of several clumps into
one, or unrealistic extensive splitting of clumps into several small
structures, or even non-detection of some structures expected to be
detected. For this reason, the initial 2D study is essential as a reference
point to understand the main structure of the cloud, which could be
significantly misrepresented by relying, uncritically and exclusively on the
3D \textsc{clumpfind} analysis.  The best configuration parameters we found
for this analysis were: the first contour level, $\mathrm{T_{low}}$, of 0.6~K;
the global noise level of the data, r.m.s., of 0.2~K; and the spacing between
the contour levels, $\mathrm{\Delta T}$, of 0.05~K.

\begin{figure}[!ht]	
\centering
	\includegraphics[width=0.48\textwidth]{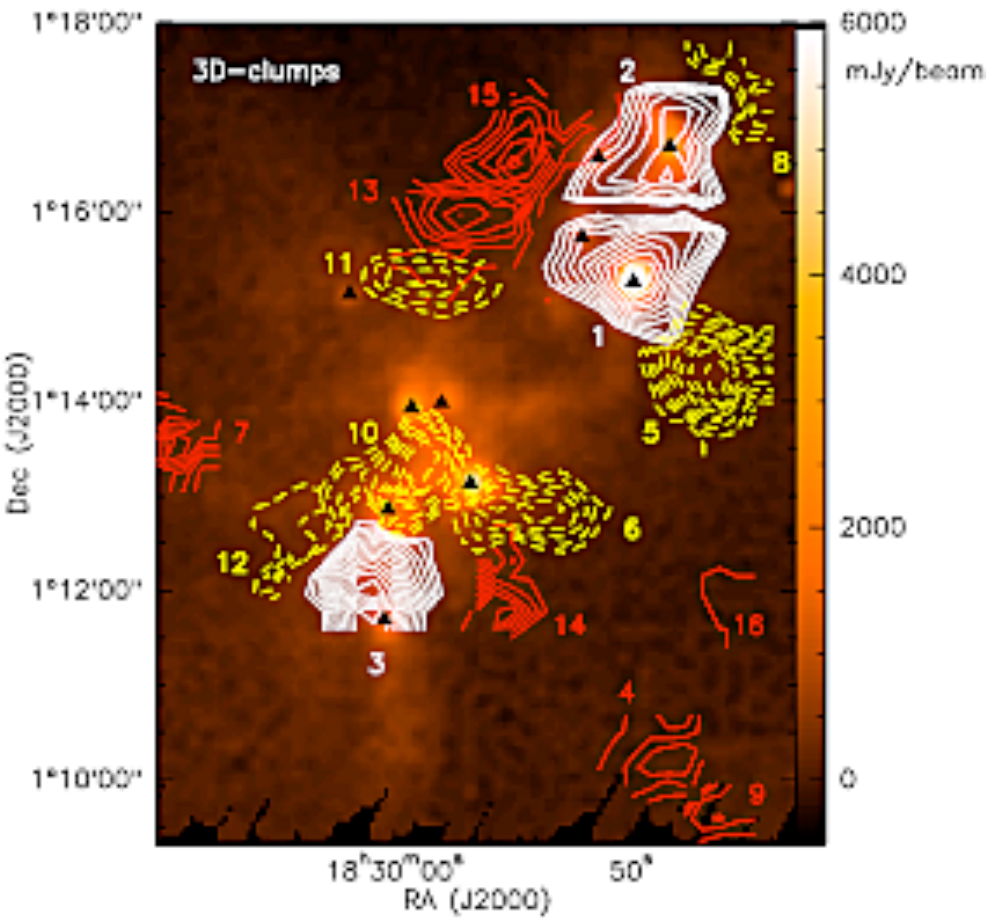}
	\caption{\small{SCUBA map of the 850 $\mu$m continuum emission (colour
            scale) overplotted with all the C$^{17}$O J=1-0 3D-clumps
            integrated intensity maps (contours and numbers).  {The numbering
              of the cores is based on their peak intensity.}  The different
            contour style and colours identify clumps starting at different
            contour levels. The white solid contours are the clumps with the
            stronger {integrated} emission, with contours starting at
            1~K\,kms$^{-1}$, the dashed yellow contours start at
            0.3~K\,kms$^{-1}$, and the solid red contours start at
            0.1~K\,kms$^{-1}$ step. For all clumps, the contour step is
            0.1~K\,kms$^{-1}$.}}
	\label{fig:3D-clumps}
\end{figure}

This analysis identified a total of 16 clumps which will be called the
3D-clumps hereafter.
These clumps are shown on Figure~\ref{fig:3D-clumps} as integrated intensity
maps in T$_{\mathrm{A}}^{*}$.kms$^{-1}$ plotted over the continuum 850~$\mu$m
data from SCUBA.  Table \ref{tab:3D-clumps} shows the properties of the 3D
clumps as numbered and plotted on Figure \ref{fig:3D-clumps}.  The nine first
columns are as in Table~\ref{tab:2D-clumps} {the last column being the
  intensity at the peak position in T$_{\mathrm{A}}^{*}$.} Once again, masses
were calculated after correcting for the IRAM 30m telescope efficiency for the
C$^{17}$O J=1$\rightarrow$0.

{Due to the difficulty in interpreting partial spectra split by 3D
  \textsc{clumpfind} between multiple spatially coincident clumps}, the mean
line width of the 3D-clumps was recovered using a different approach to that
used for the 2D-clumps.  The velocity dispersion, $\sigma$, of each clump was
estimated by {determining the velocity range where the emission of the clump
  was above $e^{-1/2}$ of its peak intensity. This was done by visually
  inspecting these thresholded channel maps of each clump.}  The quoted FWHM
is $2.35~\sigma$ and has an estimated uncertainty of 0.1~kms$^{-1}$, twice the
uncertainty of the peak velocity, 0.05~kms$^{-1}$.

The mass of the clumps within the NW sub-cluster inferred from the size and integrated intensity
of 3D-clumps 1, 2, 5 and 8 correspond to about 65\% of the total mass of
that sub-cluster. Including clumps number 11, 13 and 15 in this calculation,
the fraction of gas-mass in the clumps rises to 70\%.  The 3D-clumps 3, 6, 10,
12 and 14 constitute 40\% of the mass of the SE sub-cluster.  This is
consistent with the results from the 2D-clumps.

Since the velocity structure in the region can affect the deduced
clump structure, we also experimented with 3D \textsc{clumpfind} on the
C$^{18}$O J=1$\rightarrow$0 data.  The results from this differed from those
of C$^{17}$O only in that two clumps (3D-clumps 3 and 10) were subdivided
into 2 and 3 sub-clumps respectively. Collectively, these sub-clumps had
properties very similar to their respective C$^{17}$O clumps. 
The presence of these possible sub-clumps does not significantly 
alter the interpretation of the region for the purpose 
of our analysis, indicating that the C$^{17}$O clumps adequately describe Serpens.

\subsection{Structure of the region: combining information from 2D and
  3D-clumps}

The north region has two clear clumps unequivocally identified in both 2D and
3D methods: 2D-clumps A and B, which correspond to 3D-clumps 1 and 2
respectively. Both peak close to the position of the strongest submillimetre
sources in this region (SMM1 and SMM9), and trace the gas around them in good
agreement to the cold dense dust traced by the 850~$\mu$m emission. 

A region with higher velocity gas was detected with the 2D analysis as
2D-clump C which corresponds to 3D-clump 5. This region has quite strong
integrated emission making it detectable in the 2D search. However, as it
peaks at a very similar velocity to clump 1/A, the 3D search failed to
separate these two in some of our trial runs of the 3D analysis.  This clump
is associated with very little submillimetre continuum emission but quite
strong C$^{17}$O (and C$^{18}$O) emission. The fact that it is also seen in
N$_{2}$H$^{+}$ \citep[][]{2002A&A...392.1053O} and not in $^{12}$CO tracing
outflows (SFLPaper), is consistent with the possibility of this being a denser
region, close to being bound, directly associated with the NW
sub-cluster. It could, for example, be a very young prestellar core about to
become gravitationally unstable and collapse \citep{2007ApJ...655..958W}.

A region detected with the 3D analysis which was not seen in the 2D search was
3D-clump 8. This clump is detected at high velocities (8.8~kms$^{-1}$) and
seems to surround the clump 2/B associated with SMM9, perhaps as a
shell. {Although apparently somewhat super-virial clump, if affected by
  a depletion of C$^{18}$O by a factor of 2-3 (Appendix~~\ref{intensities}),
  this clump could be gas undergoing gravitational collapse.}

Finally there is also a low velocity region situated at the left of the main
clumps of the NW sub-cluster - detected as a single clump with the 2D method
(clump D, on Fig. \ref{fig:2D-clumps}) and as three separate clumps with the
3D \textsc{clumpfind} (11, 13 and 15 on Figure \ref{fig:3D-clumps}). This
region has a very small mass ($\sim$~1~-~2~M$_{\odot}$) and is about 5 times
super-virial. It seems to be a quiescent region at lower velocities
than the main cloud and connecting to the main cloud very close to the edge of
the NW sub-cluster as seen on dust emission.

The bulk of emission on this NW sub-cluster presents a very coherent
structure in space and velocity throughout. It does not appear to be
as filamentary as the SE sub-cluster and the emission appears confined
to relatively dense, cool compact regions.

The SE sub-cluster is quite different from the NW cluster, both in spatial
structure and velocity, even though this is not obvious from the dust
emission. {Note the higher virial ratios for the clumps within the SE
  sub-cluster, when compared to the ones in the NW, supporting once again the
  idea of more kinetic support in the south, even when the 3D
  velocity-separated clumps are considered.}  Comparing the 2D and 3D results,
shows the C$^{17}$O emission is more complex with none of the gas emission
peaks coincide with any of the compact submillimetre sources. The main peaks
of the C$^{17}$O emission in this region lie in the filament seen in dust
continuum emission, between the compact sources. The 2D-clumps E and F were
detected as 3D-clumps 10 and 3 respectively. However, the 3D search found a
more diffuse clump, 3D-clump 12, which peaks east of the filament {but with
  its edges still} overlapping spatially with 3D-clump 10 and 3, having lower
velocities than these two: 7.36~kms$^{-1}$ of clump 12, versus 7.68~kms$^{-1}$
and 8.67~kms$^{-1}$ of clump 10 and 3 respectively. {Despite being adjacent,
  and with overlapping edges, 3D-clumps 3 and 10 have a difference of
  1~kms$^{-1}$ between their peak velocities}. There is a similar velocity
difference between clump 10 and clump 6, west of the filament: 3D-clump 6,
detected in the 2D analysis as 2D-clump G, has a peak velocity of
8.51~kms$^{-1}$, $\sim$0.8~kms$^{-1}$ higher than its neighbour. These four
3D-clumps (3, 6, 10 and 12), with two sets of different peak velocities (at
$\sim$7.5~kms$^{-1}$ and $\sim$8.5~kms$^{-1}$), overlap with each other at low
intensities mainly throughout this filamentary structure of the SE
sub-cluster, even though their emission peaks are spatially offset. This also
shows that the double velocity structure in the SE sub-cluster (Section
\ref{c18odata}) is to some extent, recoverable from a single line fit using a
3D \textsc{clumpfind} analysis.

The remaining clumps detected in the SE sub-clusters trace the less dense gas
around this main filament. These were not detected in the 2D search mainly due
to their very narrow line widths, between 0.3 and 0.5~kms$^{-1}$, making them
faint in integrated intensity maps. Note that the dominant emission detected
east of the filament has lower velocities (3D-clump 7 has a peak velocity of
8.09~kms$^{-1}$), whereas the regions detected to the west have higher
velocities (3D-clumps 4, 9 and 16), with mean velocities from 8.20~kms$^{-1}$
to 8.80~kms$^{-1}$.

Globally, there appears to be a velocity gradient from east to west of nearly
1~kms$^{-1}$ over slightly more than 0.1~pc. However, this is not a smooth
gradient throughout, as in the filamentary structure there are
spatially-overlapping clumps with very different velocities.  This velocity
structure is further investigated using the C$^{18}$O lines, which are not
split by hyperfine structure in Section~\ref{c18ovel}.



\section{Dynamics of the cloud: velocity and line width}
	\label{c18o.res}

\subsection{Outflows}
\label{ss:outflows}

One important issue when studying gas dynamics in regions of active star
formation is the extent to which the line widths of molecular species are
influenced by outflows.  Using the available data, we looked for the influence
of outflows on the size scale of the cores by investigating the spectra
associated with all the submillimetre sources, looking for possible wing
emission.

Although wings on C$^{18}$O lines have proven to be able to trace outflow
interaction \citep{2002ApJ...573..699F}, in Serpens and with the 0.45K rms
noise of our dataset (Section \ref{iramobserv}) no wings were found. The lines
towards sources with known outflows are well fitted by a single Gaussian. For
example, Fig~\ref{fig:c18ospec} (top panel) shows the C$^{18}$O towards SMM1,
a source known to have an outflow, e.g. \citealt{1996ApJ...460L..45H}, is well
represented by a single Gaussian component. In the SFLPaper, we have also
searched for evidence of the influence of outflows in the C$^{18}$O emission
by comparing the C$^{18}$O J=3$\rightarrow$2 emission to the $^{12}$CO
J=3$\rightarrow$2 emission tracing the outflows. No correlation nor
anti-correlation between the C$^{18}$O emission and the outflows is found in
the region. Both these approaches lead us to conclude that the Serpens
C$^{18}$O emission is not influenced by outflows and, therefore, the velocity
components we detect in C$^{18}$O are related to the global cloud dynamics.

\subsection{Position - Velocity structure}
\label{c18ovel}

\begin{figure}[!ht]
	\centering
	\includegraphics[width=0.5\textwidth]{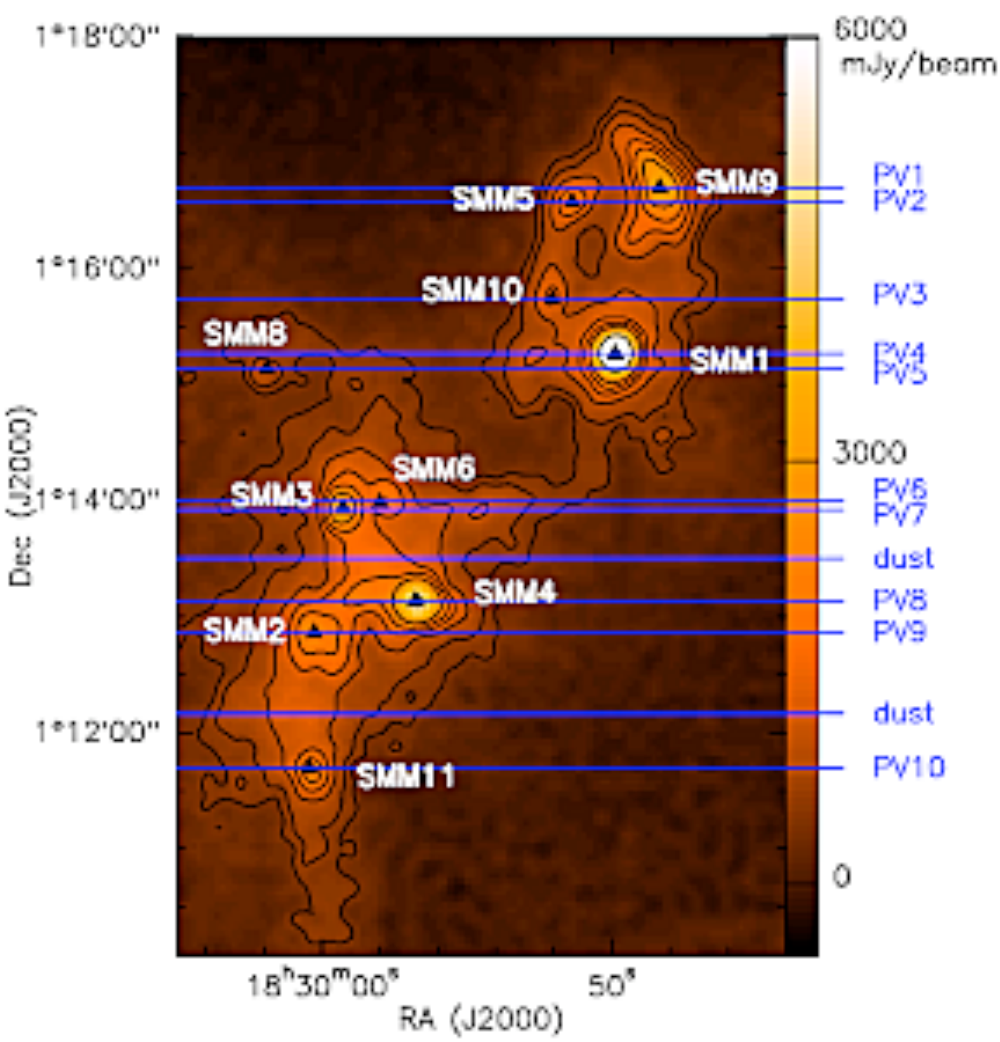}
	\caption{\small{SCUBA 850~$\mu$m map of Serpens in colour scale and
            contours (contours as in Fig.\ref{fig:srp}), showing the cuts used
            for the position-velocity diagrams of C$^{18}$O
            J=1-0 emission (Fig.~\ref{fig:northslices}, \ref{fig:southslices1}
            and \ref{fig:southslices}). The cuts made crossing the SMM sources
            are denoted as PV\#, where \# is the number of the cut,
            starting from the north. There are two cuts which are denoted
            as ``dust" which do not intercept any source, but were made to
            understand the velocity structure around the dust filament.}}
	\label{fig:cuts}
\end{figure}

\begin{figure}[!tb]
\centering
	\subfigure{PV1 \includegraphics[width=0.30\textwidth]{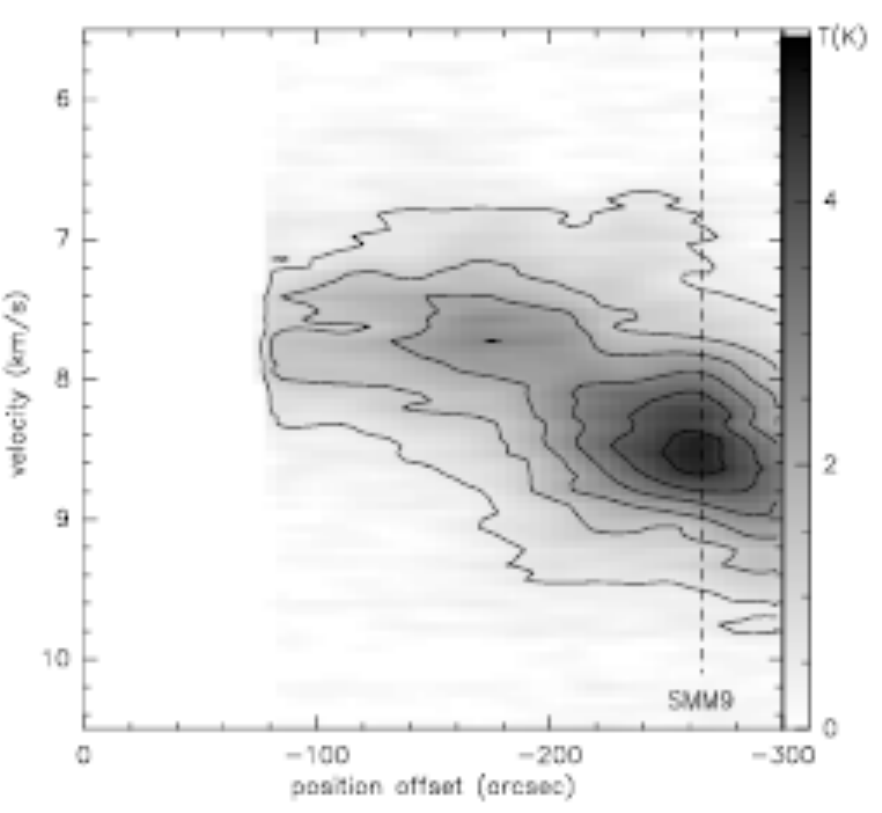}}
	\subfigure{PV2 \includegraphics[width=0.30\textwidth]{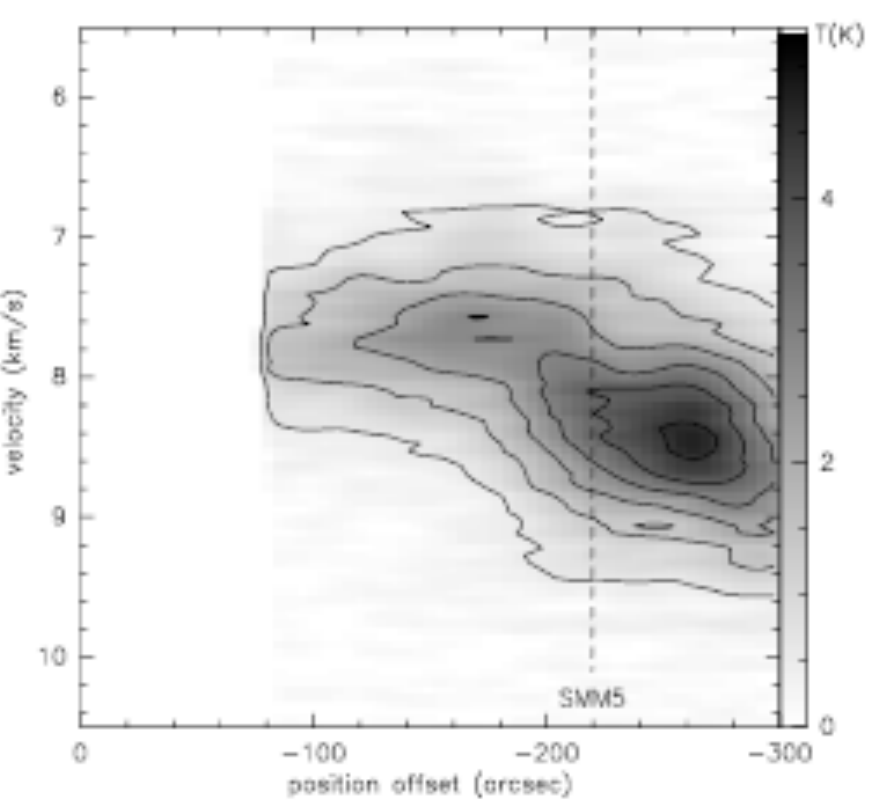}}
	\subfigure{PV3 \includegraphics[width=0.30\textwidth]{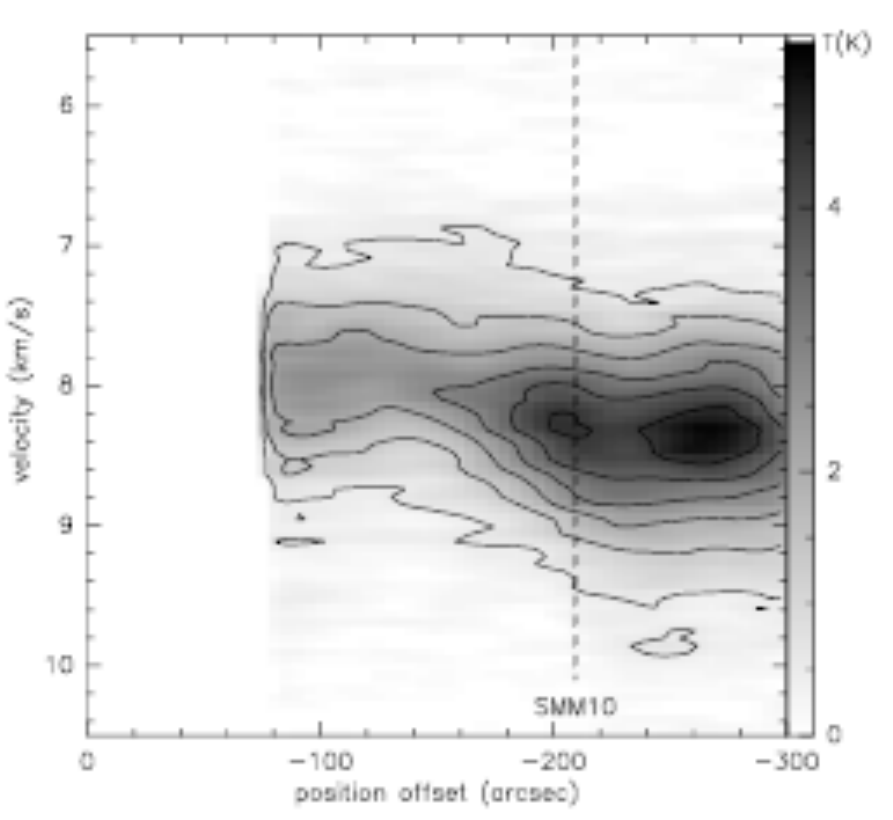}}
	\subfigure{PV4 \includegraphics[width=0.30\textwidth]{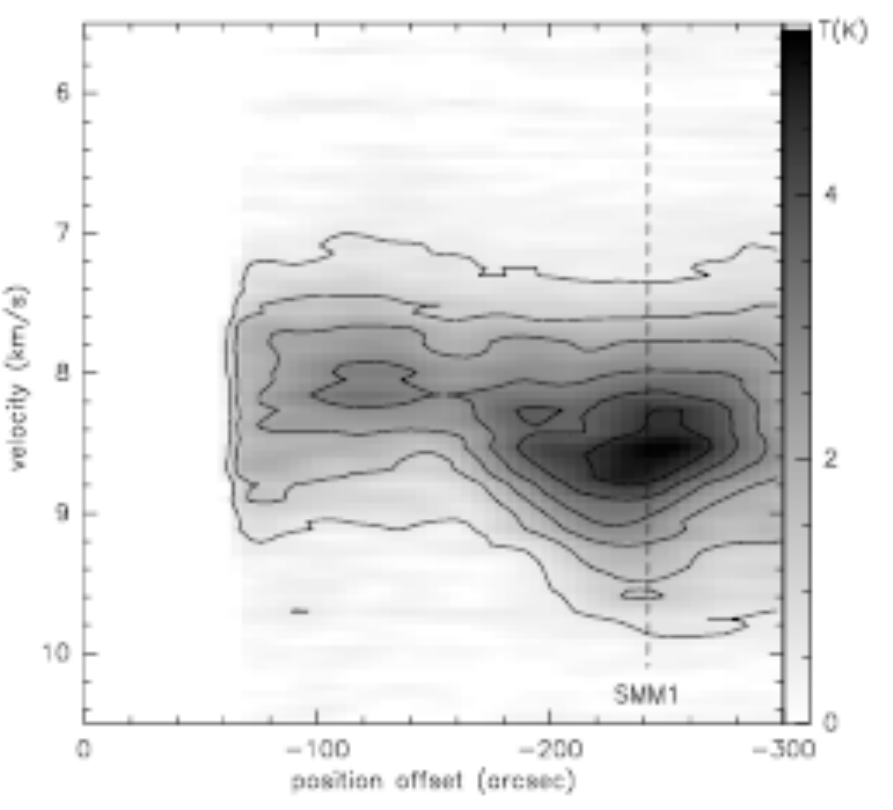}}
	\caption{\small{Position-velocity diagrams of the C$^{18}$O
            J=1$\rightarrow$0 emission. The cuts are horizontal slices of the
            map, as shown in
            Fig.~\ref{fig:cuts}. 
            The sources name and RA position are indicated in each figure
            (dashed line). The declinations are presented in
            Table~\ref{tab:SMMsources} and the RA varies from
            $18^{h}$$30^{m}$$06^{s}$ to $18^{h}$$29^{m}$$46^{s}$ (from 0 to
            -300~$''$ offset respectively). The gray scale shows the
            line intensity in T$_{\mathrm{A}}^{*}$}.}
	\label{fig:northslices}
\end{figure}

\begin{figure}[!t]
\centering
	\subfigure{PV5 \includegraphics[width=0.30\textwidth]{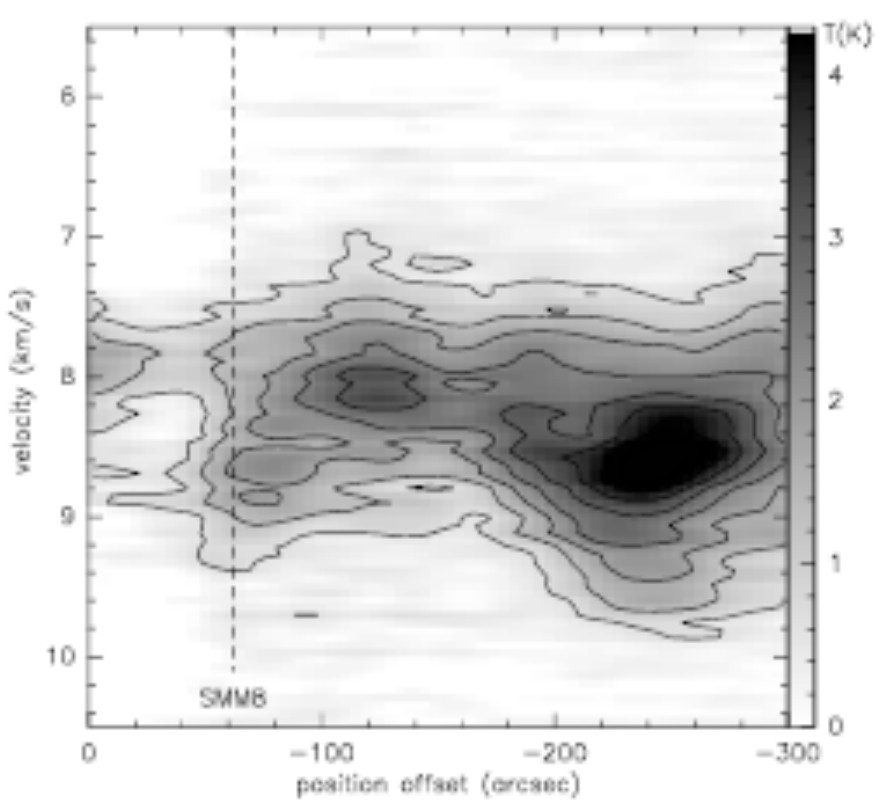}}
	\subfigure{PV6 \includegraphics[width=0.30\textwidth]{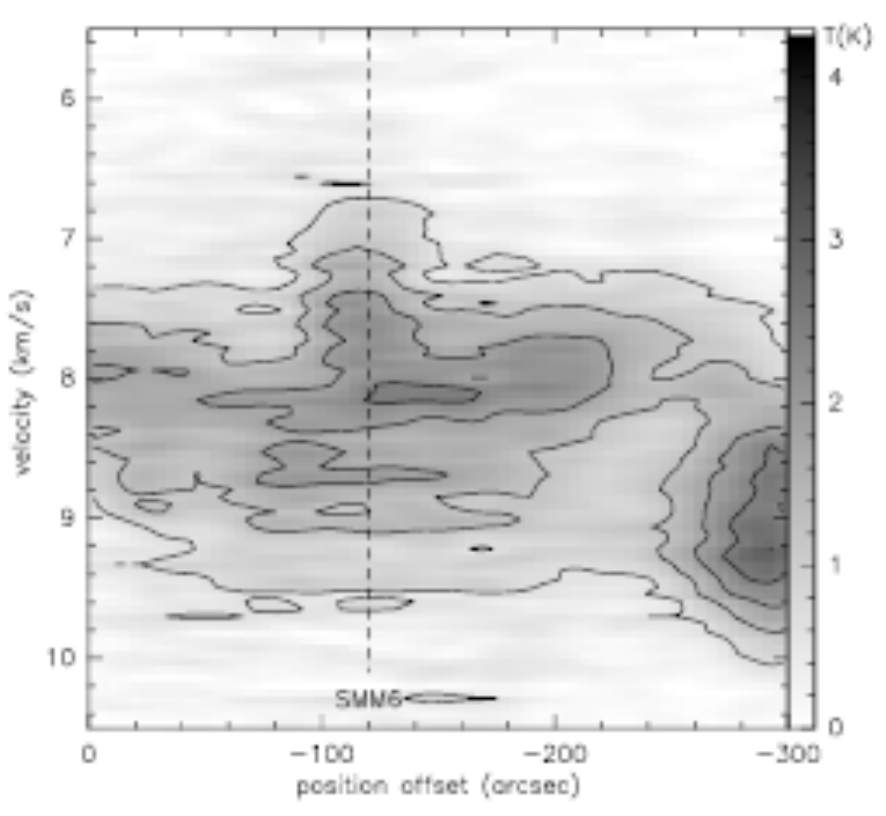}}
	\subfigure{PV7 \includegraphics[width=0.30\textwidth]{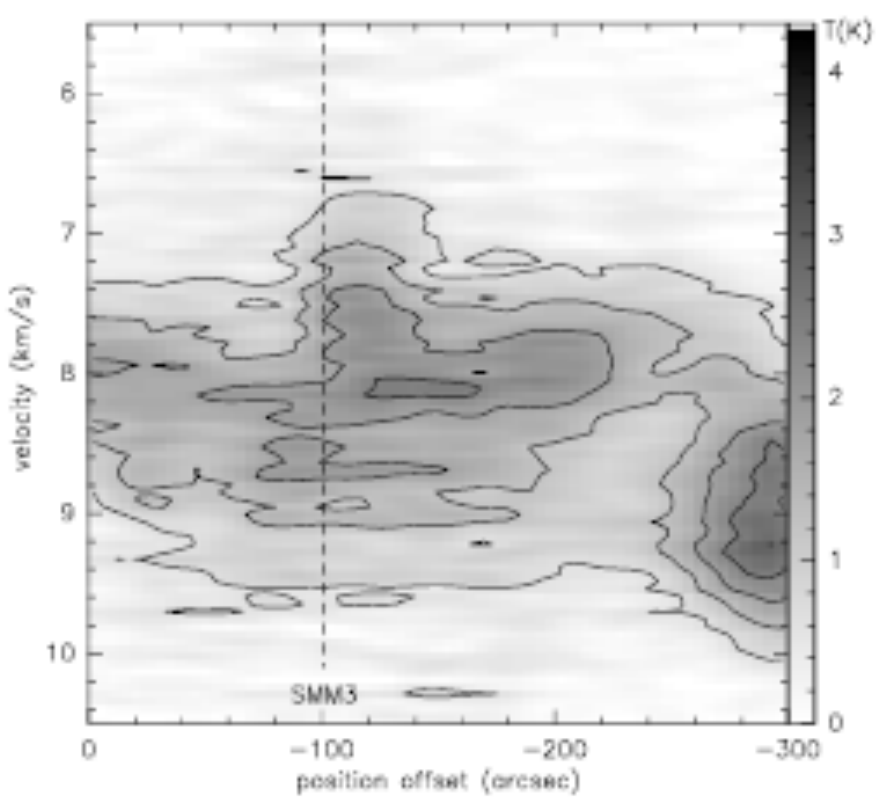}}
	\subfigure{dust \includegraphics[width=0.30\textwidth]{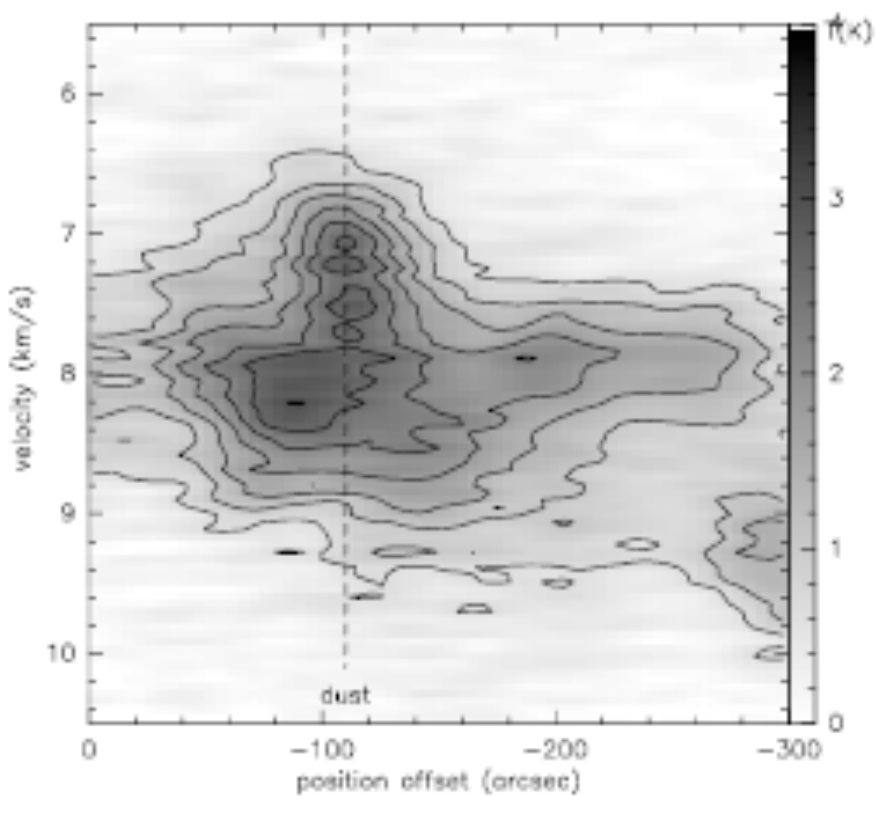}}
	\caption{\small{Same type of position-velocity diagrams as
            Fig.\ref{fig:northslices} for the SE sub-cluster. The RA also
            varies from $18^{h}$$30^{m}$$06^{s}$ to $18^{h}$$29^{m}$$46^{s}$
            (from 0 to -300~$''$ offset respectively). PV diagrams displayed
            in descending declination, as they appear on
            Fig.~\ref{fig:cuts}.}}
	\label{fig:southslices1}
\end{figure}

\begin{figure}[!t]
\centering
	\subfigure{PV8 \includegraphics[width=0.30\textwidth]{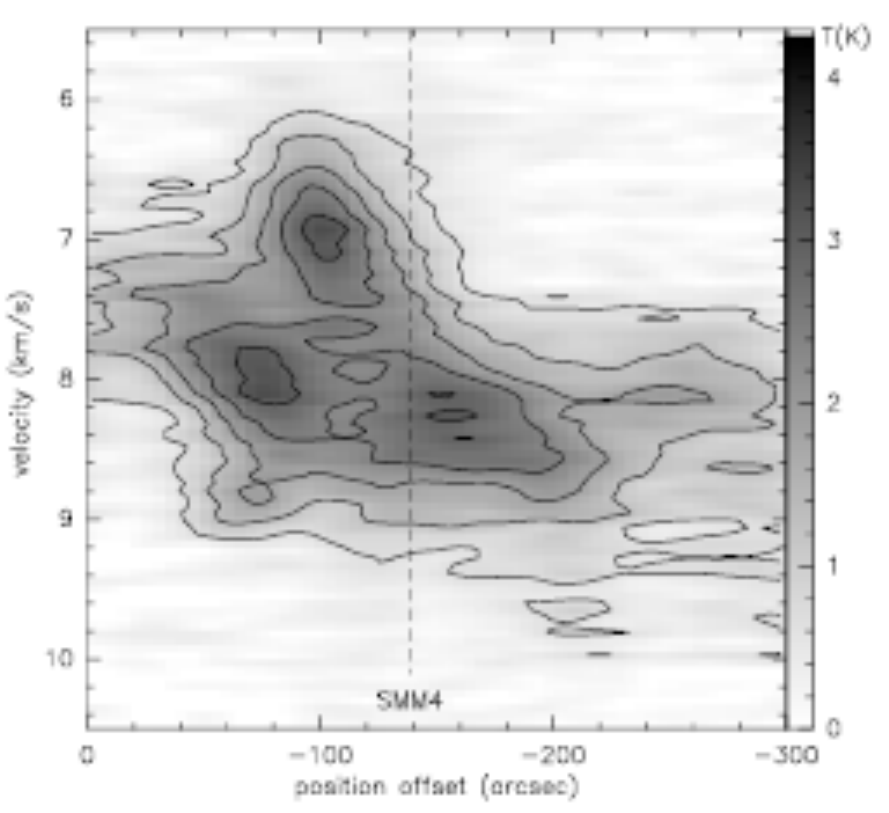}}
	\subfigure{PV9 \includegraphics[width=0.30\textwidth]{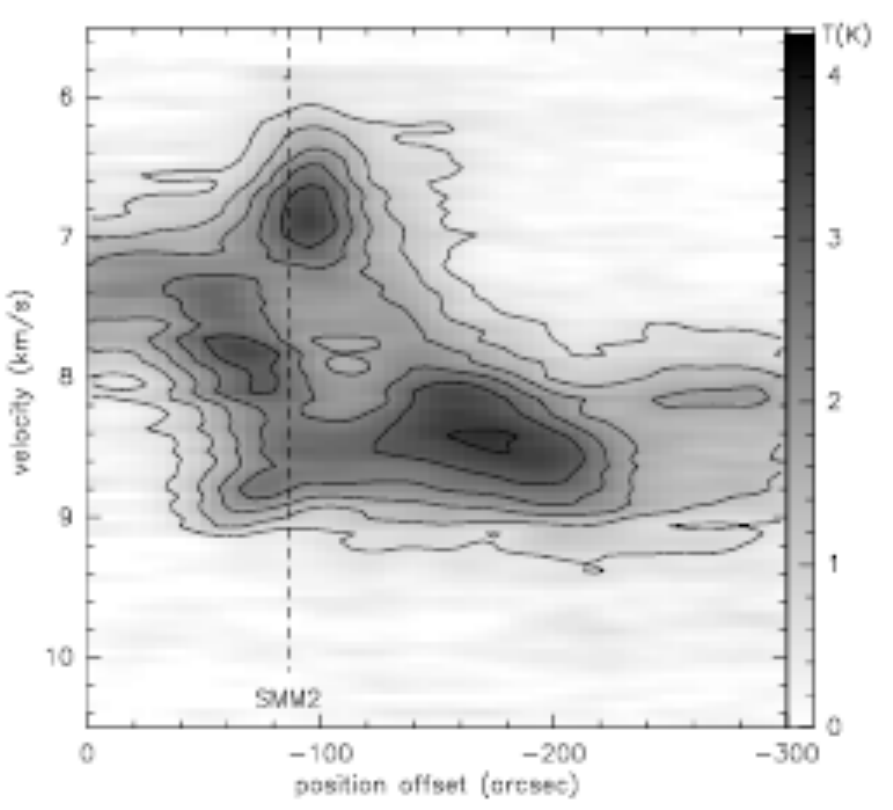}}
	\subfigure{dust \includegraphics[width=0.30\textwidth]{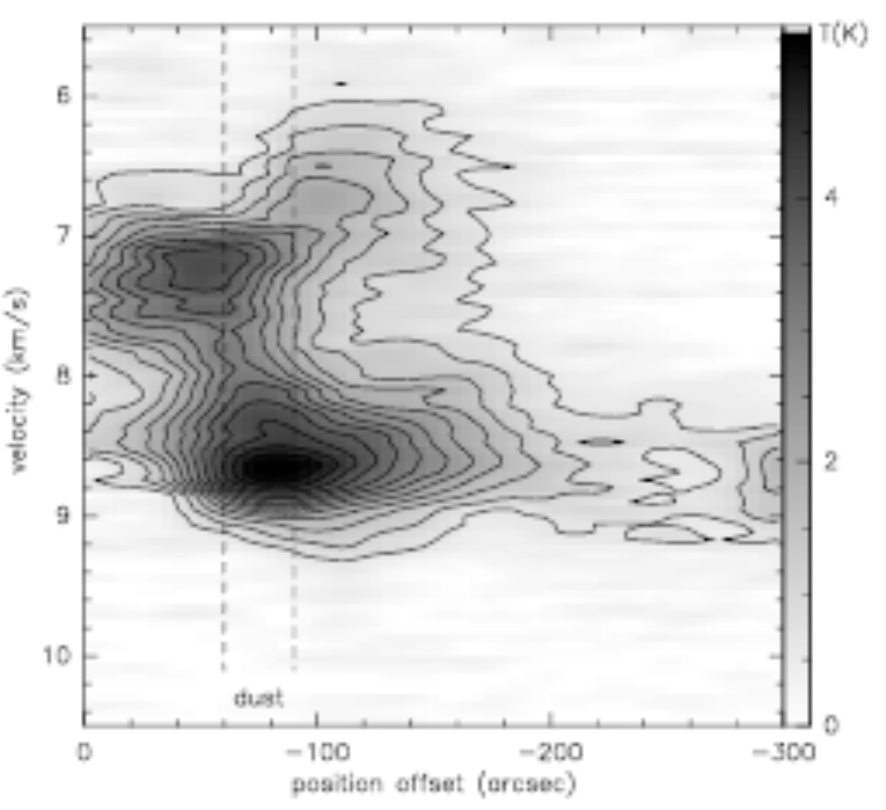}}
	\subfigure{PV10 \includegraphics[width=0.30\textwidth]{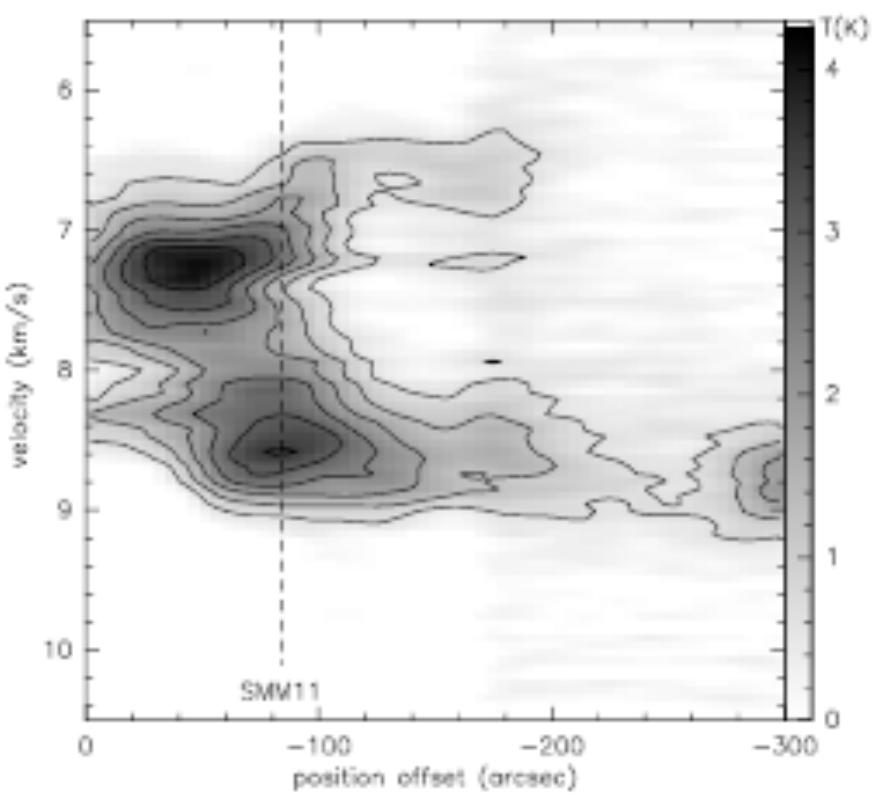}}
	\caption{\small{Fig. \ref{fig:southslices1} continued: remaining position-velocity diagrams of the SE sub-cluster as plotted on Fig.~\ref{fig:cuts}.}}
	\label{fig:southslices}
\end{figure}

As revealed by the C$^{17}$O (Section~\ref{c17o.res}), the NW region is mostly
traced by higher velocity emission, the exception being the region offset to
the east of the sub-cluster. On the other hand, the SE is not so homogeneous,
containing both higher and lower velocity components, which overlap
approximately where the filamentary structure is seen in the continuum
observations. The C$^{18}$O shows very well defined double peaked emission
along the southern region, that starts disappearing as we move
north. Figure~\ref{fig:c18ospec} shows two examples of spectra in this region,
and Fig.~\ref{fig:southslices} shows the evolution of the double component
along the map.  The existence of a double-peaked spectrum in other optically thin
tracers such as N$_{2}$H$^{+}$ \citep{2002A&A...392.1053O} rules out self
absorption as an explanation of the double peaked C$^{18}$O emission (cf
Section~\ref{c18odata}).

On the basis of a study of the line centroid velocity (despite the presence of
double peaked lines) \citet{2002A&A...392.1053O} argued that the region is
undergoing global rotation. However position-velocity diagrams of horizontal
slices along the C$^{18}$O map are incompatible with this interpretation
(Fig.~\ref{fig:cuts} and Fig.~\ref{fig:northslices} to
\ref{fig:southslices}). Moving from north to south, and slicing at the
declination of each SMM source, we can see the two separate clouds, very well
distinguished close to SMM11. For simple rotation, we would expect to see a
smooth gradient along the velocity axis as the RA changes. Instead we observe
two velocity components, clearly separated in the southern part of Serpens
(see e.g. PV10) and merging together when moving to the north of the
sub-cluster (see e.g. PV7).  At this point, the two components are barely
distinct lines, producing broad, non-gaussian profile.  Furthermore, the SMM
sources in the SE sub-cluster appear at the edges of the double velocity
region (hereafter referred to as the interface), whilst the filamentary
structure seen in dust follows the interface region itself (see the PV
diagrams labeled as ``dust" on Fig.~\ref{fig:southslices1} and
\ref{fig:southslices}).

The lower velocity cloud (hereafter LVC) appears to be interacting with the
high velocity cloud (hereafter HVC), apparently provoking the enhanced dust
emission between SMM2, SMM3, SMM4 and SMM6 - and also the elongated filament
that extends south towards SMM11 and beyond. A dynamical interaction between
two clouds, as indicated by this space-velocity structure and the turbulent
motions found towards the south, might have triggered this episode of star
formation along the filament (Sec.~\ref{discussion}).

\begin{verbatim}








\end{verbatim}


\subsection{Decomposition of the C$^{18}$O line components}
\label{sec:decomp}

\begin{figure*}[!t]
	\centering
	\hfill
	\hfill
	\includegraphics[width=0.85\textwidth]{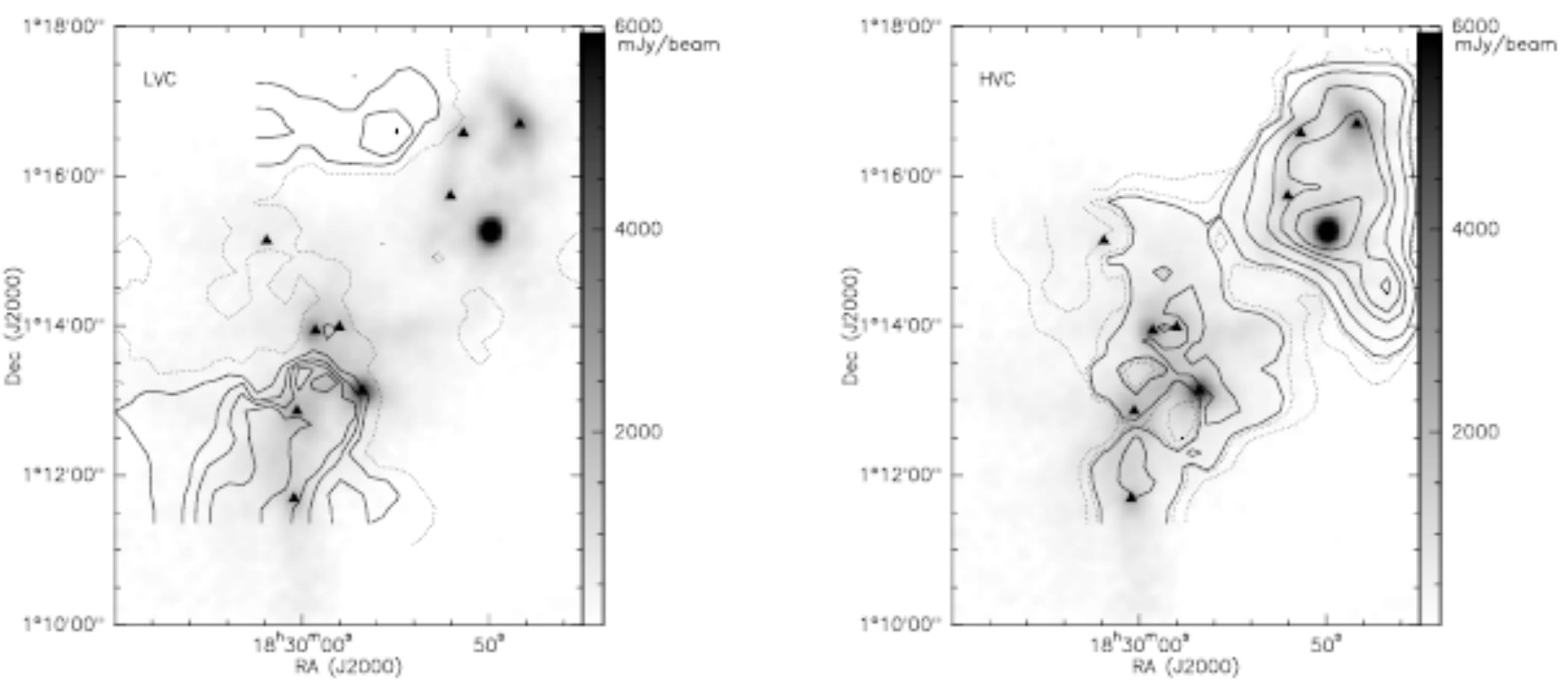}
	\hfill
	\hfill
	\caption{\small{Integrated intensity maps as a result of the
            separation of the two line components of the C$^{18}$O J=1-0
            transition, under the assumption of two different clouds seen
            along the line of sight. The LVC is shown on the left and HVC on
            the right. The background grey scale shows the 850~$\mu$m emission
            tracing the cold dust, with its respective submillimetre sources
            plotted as triangles. The contours represent the integrated
            intensity of the modelled Gaussians (fitting the data in
            T$_{\mathrm{A}}^{*}$). Contour key: left (LVC) at
            0.2~K\,kms$^{-1}$ (dashed) and 1.5, 2.5, 3.2 and 4.0~K\,kms$^{-1}$
            (solid); right (HVC) at 2.5 and 3.0~K\,kms$^{-1}$ (dashed) and 3.2,
            4.0, 5.0, 5.5 and 6.0~K\,kms$^{-1}$ (solid).}}
	\label{fig:hvc-lvc}
\end{figure*}

\begin{figure*}[!ht]
	\centering
	\hfill
	\includegraphics[width=0.80\textwidth]{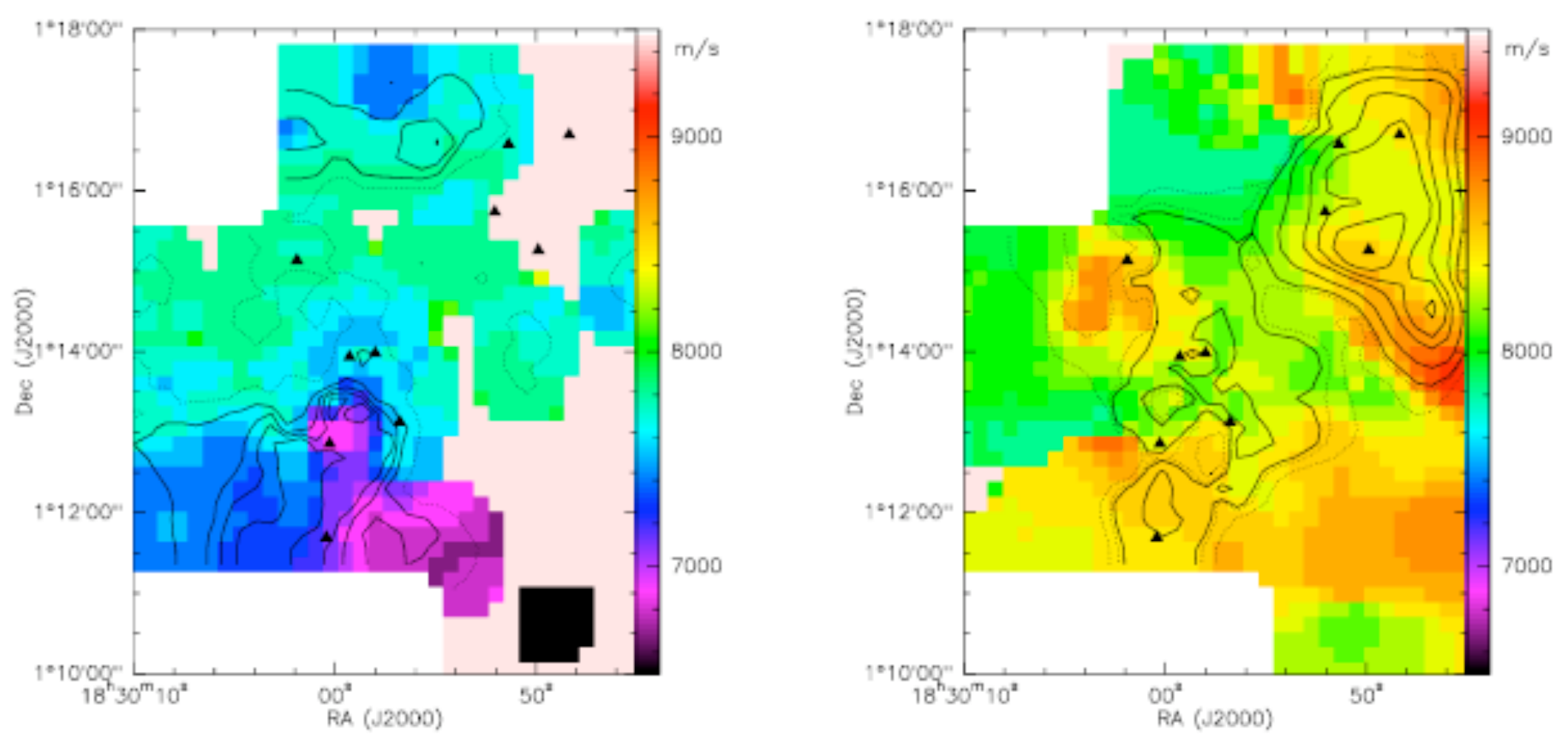}
	\hfill
	\hfill
	\caption{\small{Velocity structure maps of Serpens, as a result of the
            separation of the two line components of the C$^{18}$O J=1-0
            transition. As in Fig.~\ref{fig:hvc-lvc}, the LVC is represented
            on the left and HVC on the right. The submillimetre sources are plotted as triangles, 
            and the contours represent the integrated intensity as in Fig.~\ref{fig:hvc-lvc}. 
            The colour scale is now the centroid velocity of the same modelled Gaussians, 
           	where the light pink colour represents the lack of a fit to that velocity component.}}
	\label{fig:highlow}
\end{figure*}

To investigate the velocity structure of the C$^{18}$O J=1$\rightarrow$0
emission we have decomposed the datacube in two, by fitting two
velocity components to the C$^{18}$O spectra and then creating a 
model datacube from the Gaussian fits, for each of the two components. 
The results of this decomposition are shown in Fig.~\ref{fig:hvc-lvc} and Fig.~\ref{fig:highlow}.

The data were first rebinned to 0.1~kms$^{-1}$ velocity channels. Then, for
each spectrum, the line was fitted with a single Gaussian and then a double
Gaussian. The two Gaussian fit was selected as the model for the line only if:
{\it i)} the difference between the central velocities of the two Gaussian fit
($\Delta V$) was greater than 0.35~kms$^{-1}$ 
or {\it ii)} both lines were relatively strong with the peak intensity ratio
of the stronger to the weaker line less than 2.4.  The value of 2.4 was
determined by a careful analysis of various line fits which showed that if the
ratio was more than 2.4, the weaker line fit was poorly fit.  The remaining
spectra were fitted with a single Gaussian. An example of this fitting is
shown on Figure~\ref{fig:c18ospec} (lower panel). From this fitting procedure,
two model datacubes were created, one for each velocity component, allowing us
to study the two clouds separately. The higher velocity component (HVC) of the
double peaked lines, as well as the single lines with central velocity greater
than 7.8~kms$^{-1}$, were included in the HVC datacube; lower-velocity lines
and single lines peaking below 7.8~kms$^{-1}$ were incorporated in the LVC
model datacube.

Figure~\ref{fig:hvc-lvc} shows the spatial distribution of the LVC and HVC
using the integrated intensity from the model datacubes. Overall, the HVC
traces the distribution of the 850~$\mu$m continuum emission better than the
LVC. The HVC emission is stronger in the north, but it lies along the filament
containing both sub-clusters, extending roughly in a SE-NW direction.  The LVC
is roughly aligned along the S-N direction and is stronger in the south, where
it meets the HVC.

Figure~\ref{fig:highlow} shows again the integrated intensity of the 
modelled datacubes for each component, but shows also the velocity structure of 
each cloud. In the NW sub-cluster, the HVC appears at velocities around 8.4~kms$^{-1}$, 
with the exception of a few regions at the edges of the cloud, which appear to reach
velocities as high as 8.8~kms$^{-1}$. The region which stands out from the
bulk of this sub-cluster is the region SW of SMM1, which is not present in the
dust emission even though it is rather strong in gas emission, having the
highest velocities of the entire cloud (reaching 9~kms$^{-1}$).
The LVC, in the NW sub-cluster, is spatially offset to east, with velocities 
of 7.5~-~7.8~kms$^{-1}$, similar to most of the emission in the south.

The region between the two sub-clusters, dominated by the emission from the
HVC, has the systemic velocity of Serpens (around 8.0~kms$^{-1}$) possibly due
to the merging of the two components. Note that the emission here is rather
weak, and the presence of SVS2, a more evolved (flat spectrum) near-IR source
\citep[][]{2004A&A...421..623K}, suggests this region may be more evolved.

In the SE sub-cluster, the HVC velocities range from 8 and 8.5~kms$^{-1}$, 
being higher towards the southern end of the filament. 
On the other hand, the LVC shows a velocity gradient increasing 
from west to east - contrary to the HVC. The material west of the southern 
filament, has velocities of about 6.8~-~7~kms$^{-1}$. At the centre of the 
filament the velocities are around 7.5~kms$^{-1}$, translating into a gradient 
of $\sim$~5~kms$^{-1}$pc$^{-1}$. To the east of the filament the velocities 
are approximately constant and around 7.5~kms$^{-1}$. 
Therefore, it seems that the clouds have a greater offset in velocities 
in the far-south end of the filament, converging into one
intermediate velocity as one moves north. When two lines can no longer be
separated, the emission becomes a single broader line, centered at the
intermediate velocities ($\sim$~8~kms$^{-1}$).

The line width in the SE sub-cluster, specially where the two components
merge, is around 2~kms$^{-1}$. This is almost twice the line width of the NW
($\sim$~1~kms$^{-1}$). This difference is reflected as a four times higher
kinetic support in the SE region, in comparison to the NW region. This is
consistent with the C$^{17}$O J=1$\rightarrow$0 analysis (Section
\ref{c17o.res}), where it was showed that the NW sub-cluster is a bound
structure, whereas the SE sub-cluster was somewhat super-virial.  The SE is
therefore much more dynamic than the NW, as already foreseen by the C$^{17}$O analysis.


\section{Physical properties: Temperatures and column densities}
\label{scatter}

The optical depth of the C$^{18}$O can be estimated from the ratio of the
integrated intensities of the C$^{18}$O and C$^{17}$O J=1$\rightarrow$0
transitions.  Over the mapped region the observed ratio shows little coherent
spatial structure and is approximately constant with a value consistent with
the abundance ratio of the species, $\sim$3.5
\citep[e.g.][]{1980IAUS...87..397P,1982ApJ...262..590F} implying the emission
from both species is optically thin.

\begin{figure}[!t]
  \centering
	\includegraphics[width=0.45\textwidth]{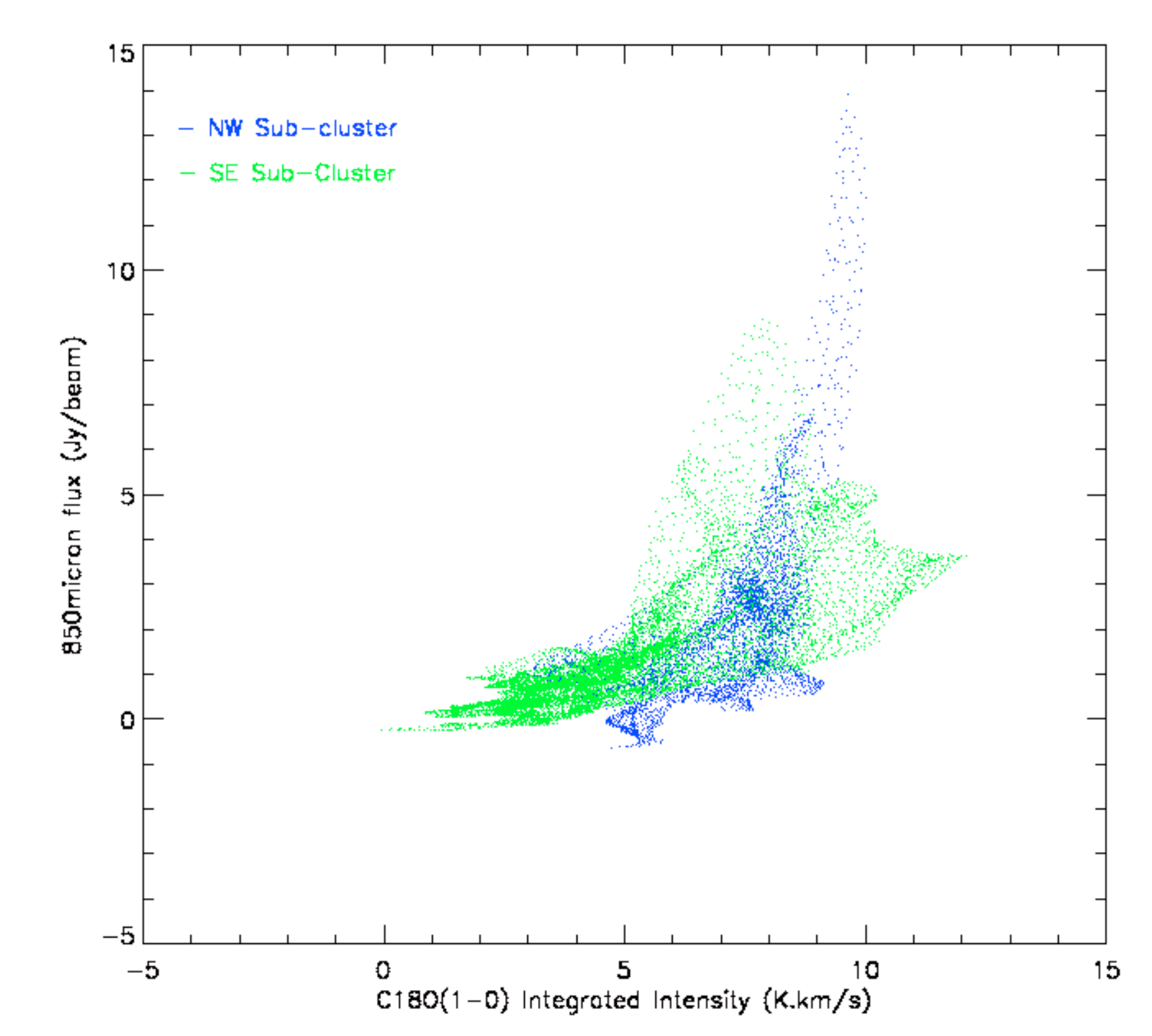}
	\hfill
	\includegraphics[width=0.45\textwidth]{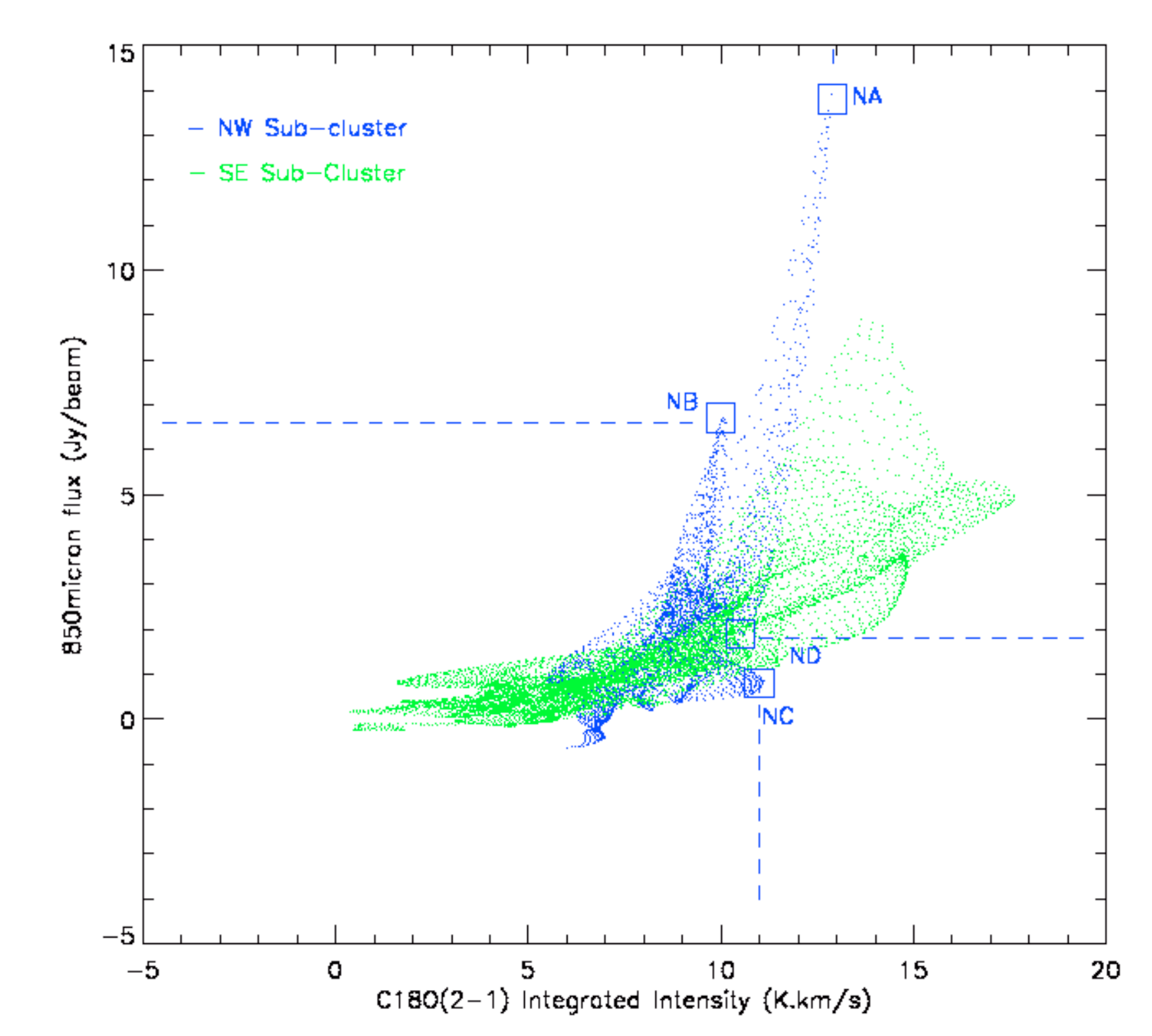}
	\hfill
	\includegraphics[width=0.45\textwidth]{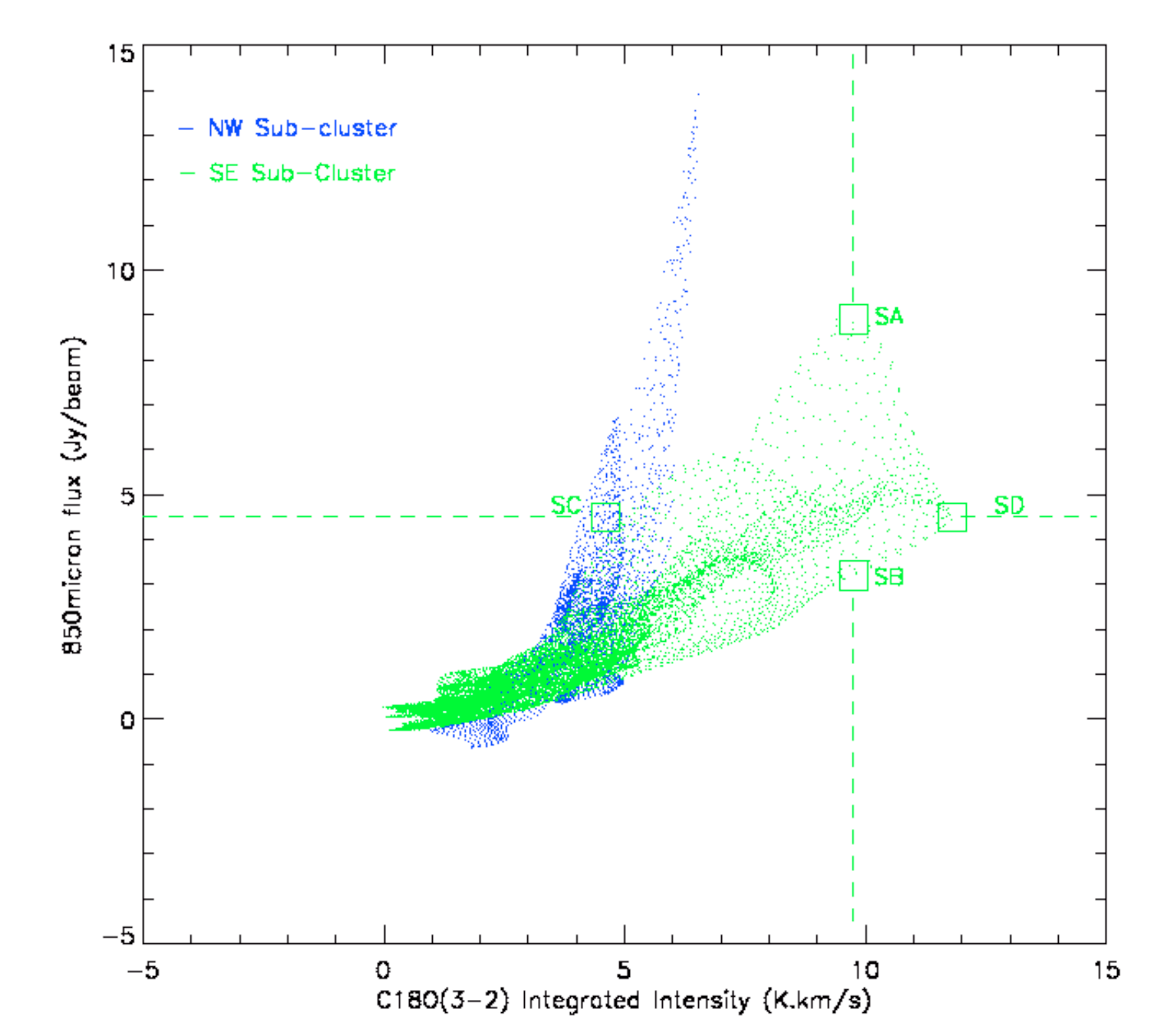}
        \caption{\small{Scatter plots of the SCUBA 850~$\mu$m dust emission
            against the C$^{18}$O integrated intensity over the whole range of
            velocities (top: J=1$\rightarrow$0, middle:
            J=2$\rightarrow$1, bottom: J=3$\rightarrow$2).  Blue points show
            the NW sub-cluster and green points the SE sub-cluster.  The four
            positions chosen in each sub-cluster to investigate with non-LTE
            modelling shown on the middle panel for the NW region and the
            lower panel for the SE region. These positions are also indicated
            on Fig.~\ref{fig:mappositions}.}}
  \label{fig:scatter}
\end{figure}

To better understand the correlation between the dust and gas in this region,
Figure~\ref{fig:scatter} shows a pixel-by-pixel comparison of the 850~$\mu$m
flux density against the integrated intensity of the three transitions of
C$^{18}$O, all convolved to a common resolution of 24$''$. {For the purpose of
  these scatter plots, we have oversampled the data to a pixel size of
  $2.5''$, in order to better distinguish the trends.}

\begin{figure}[!ht]
  \centering
	\includegraphics[width=0.47\textwidth]{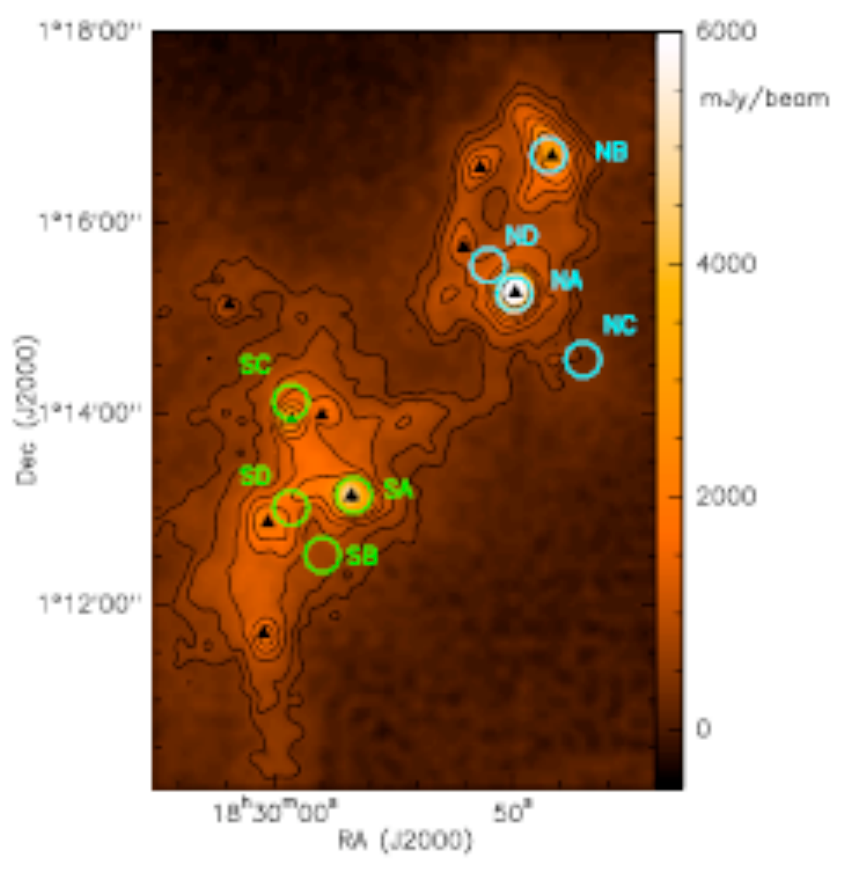}
  \hfill
  \caption{\small{Map showing the positions of the selected regions for the
      non-LTE RADEX study indicated by blue and green circles (for the
      positions in the NW and SE sub-clusters, respectively) and labeled as in
      Fig.~\ref{fig:scatter}. The contours and colour scale show the SCUBA
      850~$\mu$m emission as in Fig.~\ref{fig:cuts}.}}
  \label{fig:mappositions}
\end{figure}

Overall, the distribution of points is very similar for the three
transitions. There is a general correlation between dust and gas, especially
for the weaker emission (Fig.~\ref{fig:scatter}).  However, the distributions
also show structure which consistently appears across all three
transitions. The very prominent peaks of dust emission corresponding to the
stronger submillimetre sources are obvious, and although in general there is
an increase in the C$^{18}$O emission at these positions, the dust peaks do
not correspond to global peaks in the C$^{18}$O emission. Indeed the nature of
the relationship between the C$^{18}$O emission and the dust appears different
in the NW and SE sub-clusters.

Focusing on the NW region (blue in Fig.~\ref{fig:scatter}), the plots are
dominated by the two dust peaks, each of which is associated with a well
defined, but separate, increase in C$^{18}$O emission.  Comparing the
C$^{18}$O intensity, the emission becomes weaker moving to higher energy
transitions.  On the other hand, the SE sub-cluster (green in the figure)
shows a different trend from transition to transition, becoming stronger at
higher transitions. In addition, there appears to be a more pronounced general
correlation in this region between the dust and line emission.  Nevertheless,
there are clearly structures departing from this trend: several 850~$\mu$m
peaks corresponding to SMM sources; and C$^{18}$O peaks, which do not have
significant submillimetre emission.

\subsection{LTE Analysis}
\label{rotdiag}

From the dust continuum emission, the volume densities in the Serpens
sub-clusters are typically higher than the critical densities of each of the
three transitions (Table~\ref{tab:criticdens}) observed here. We therefore
initially calculate the gas properties assuming LTE (local thermodynamic
equilibrium) using a rotation diagram analysis. Despite its uncertainties, the
rotation diagram method is robust in retrieving the column density structure
and trends throughout the region, as well as the approximate absolute column
densities.

\begin{figure}[!t]
  \centering
	\includegraphics[width=0.44\textwidth]{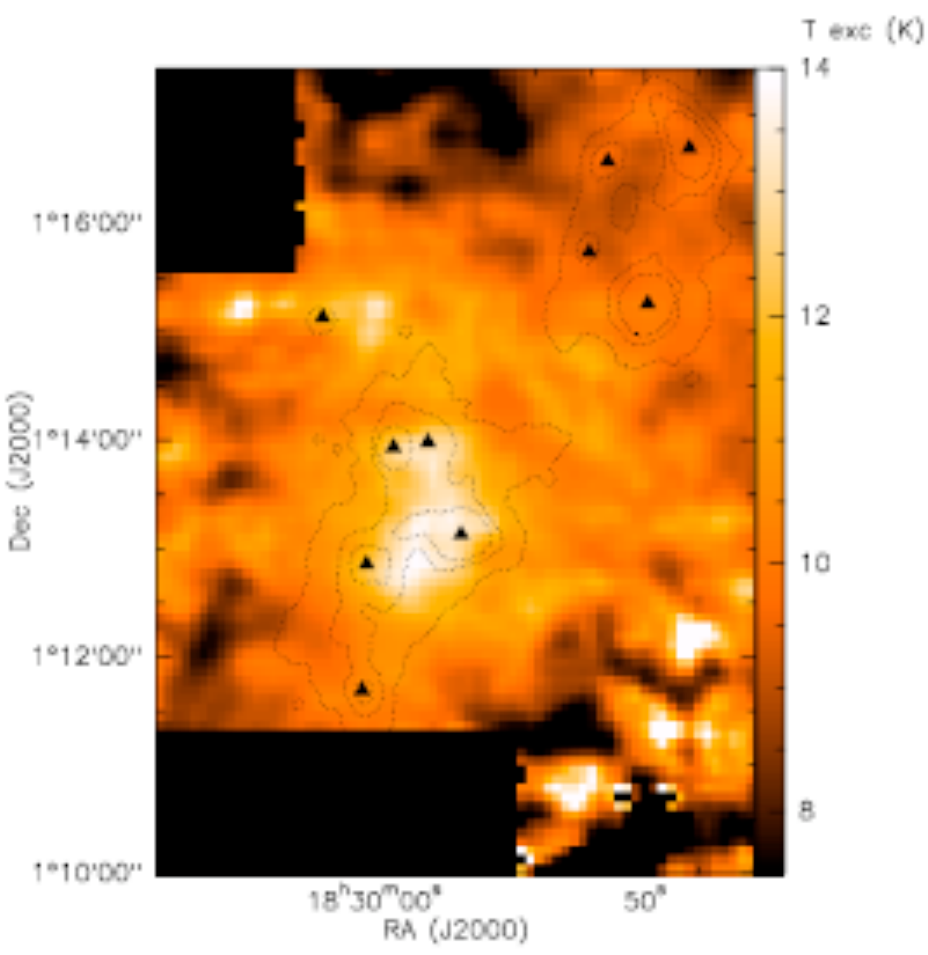}
	\hfill
	\hfill
        \caption{\small{LTE excitation temperature map in colour scale.  The
            dotted black contours show the dust 850~$\mu$m emission at 0.6,
            1.2 and 1.8~Jy~beam$^{-1}$.}}
  \label{fig:exc}
\end{figure}

\begin{figure}[!t]
  \centering
	\includegraphics[width=0.48\textwidth]{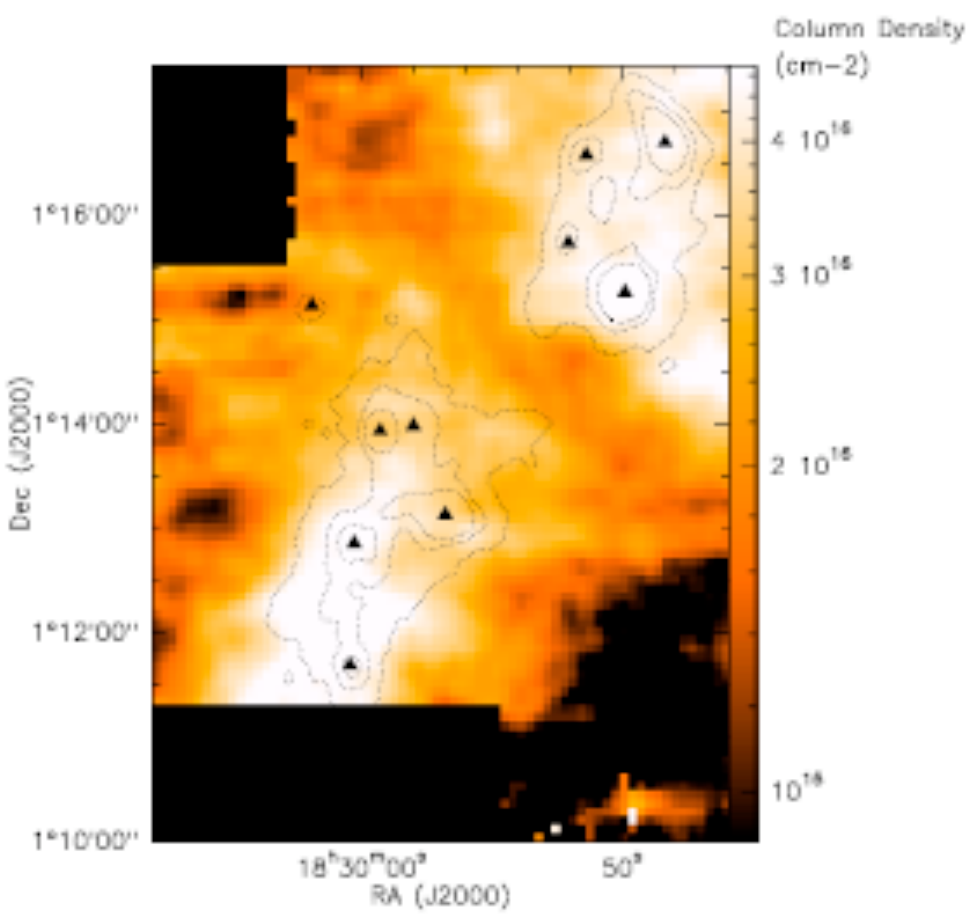}
	     \caption{\small{Column density map (colour scale) derived from the
            rotation diagram method. The dust 850~$\mu$m emission is
            overplotted in dotted black contours, at the same levels as in
            Fig.~\ref{fig:exc}.}}
  \label{fig:columndens}
\end{figure}

\begin{table*}[ht]
	\caption{Critical densities (n$_{\rm critical}$) at 10K and 20K for 
C$^{18}$O}
		\centering
		\begin{tabular}{c c c c }
		\hline 
		\hline
		C$^{18}$O 	& A$_{\mathrm{u}}$ 	 & K$_{\mathrm{u}}$					  		 			& n$_{\mathrm{critical}}$  \\
		transition  & (s$^{-1}$) & (10$^{-11}$~cm$^3$s$^{-1}$)  & (10$^{3}$~cm$^{-3}$) \\
								&						 & 10K -- 20K											&	10K -- 20K \\	
		\hline
		J=1$\rightarrow$0	& 6.266$\times$10$^{-8}$ & 3.3 -- 3.3  & 1.9 -- 1.9  \\  	 
		J=2$\rightarrow$1	& 6.011$\times$10$^{-7}$ & 7.2 -- 6.5  & 8.3 -- 9.3  \\    
		J=3$\rightarrow$2	& 2.172$\times$10$^{-6}$ & 7.9 -- 7.1  & 27 -- 30  \\  	 
		\hline
		\end{tabular}
		\begin{flushleft}
		{These values were derived using the information provided by the Leiden Atomic and Molecular
     Database\footnotemark \ \citep{2005yCat..34320369S}: A$_{\mathrm{u}}$ being the Einstein Coefficient for the upper level, and K$_{\mathrm{u}}$ the respective collision rate at both 10~K and 20~K.}
    \end{flushleft}
		\label{tab:criticdens}
\end{table*}

Fig.~\ref{fig:exc} shows a map of the excitation temperature across the region
constructed from the 24$''$ resolution integrated intensity maps sampled with
5$''$ pixels (as in the original IRAM data).  This shows the NW and SE
sub-clusters to have different temperature structures. The NW appears very
homogeneous with no significant temperature peaks and with temperatures
ranging from 9 to 10~K.  In contrast the SE region has both higher
temperatures, ranging from $\sim$~10 to 14~K and a much more peaked
distribution. Interestingly, this enhanced temperature in the south does not
peak on the SMM protostars but rather between them, along the dust filament
which corresponds to the interface region seen on the PV diagrams
(Fig.~\ref{fig:southslices1} and \ref{fig:southslices}).

The C$^{18}$O column density map (Fig.~\ref{fig:columndens}) calculated from
the rotation diagram recovers more of the dust structure than the temperature
map. Both the south and north sub-clusters are evident as denser regions, even
though the dust and gas column densities peaks are not always coincident,
especially in the SE. The mean C$^{18}$O column densities are very similar in
the north and the south. The regions with higher gas column density (the
entire NW sub-cluster and the filament between SMM11 and SMM2 in the SE
sub-cluster) have a lower temperature. Conversely the regions with slightly
lower gas column density (between SMM2, SMM4, SMM3 and SMM6 in the SE
sub-cluster) have higher temperature.  The region south-west of the NW
sub-cluster which appears to have a relatively high gas column density seems
to have very similar properties to the rest of the NW sub-cluster and yet it
is not detected in dust emission.

\footnotetext[3]{http://www.strw.leidenuniv.nl/moldata/}

\subsection{Non-LTE Analysis}
\label{radex.results}

To better understand the physical conditions of Serpens and the apparent
discrepancies between the gas and dust emission, we have selected 4 key
positions in the NW and another 4 in the SE using the scatter plots
(Fig.~\ref{fig:scatter}) for more detailed analysis.  The selected positions
are indicated on Fig.~\ref{fig:mappositions}.  These positions correspond to
interesting features in the correlations between the dust and gas emission,
selected to span the range of the correlation.  For these positions we
performed a non-LTE analysis using RADEX\footnote{RADEX is a statistical
  equilibrium radiative transfer code, available as part of the Leiden Atomic
  and Molecular Database. The formalism adopted in RADEX is summarized in
  \cite{2007A&A...468..627V}}.

We created a $500\times500$ grid of gas column density (ranging from 10$^{12}$
to 10$^{19}$~cm$^{-2}$) and temperature (ranging from 5~K to 40~K). For each
grid point we used RADEX to calculate the C$^{18}$O integrated intensities for
all three transitions (denoted as $I_{(J_{up}-J_{low})}$).  For each position
modelled, a volume density determined from the submillimetre dust continuum
emission was adopted, assuming a cloud depth of 0.2~pc (based on the projected
size of the dust emission), a dust opacity of 0.02~cm$^{2}$g$^{-1}$ at
850~$\mu$m \citep[][]{1999ApJ...522..991V,2006ApJ...653..383J} and a dust
temperature of 10~K for all but three positions. The three exceptions are: {
  position NA ($\equiv$SMM1) where 38~K was adopted from the SED fit by
  \citet{1999MNRAS.309..141D}, and positions NB($\equiv$SMM9) and
  SA$(\equiv$SMM4), where we adopted a temperature of 25~K, consistent with
  the $>20$~K determined by \citet{1999MNRAS.309..141D}.  For each of the
  southern positions, we assumed the H$_{2}$ volume density to be the same for
  both C$^{18}$O velocity components. Changing the dust temperature or the
  assumed cloud depth changes the estimated the H$_{2}$ volume densities,
  however this only becomes important if the derived volume densities becomes
  lower than the critical densities for our transitions.  We have tested these
  effects using RADEX, and the resulting column densities and kinetic
  temperatures remain unaffected by changes in the assumed dust temperature
  between 10~K and 40~K, or in the assumed depth between 0.1pc and 0.3pc.  If
  the transitions are thermalised, only the fractional abundance of C$^{18}$O
  will be affected by changing the assumed H$_{2}$ column density.}

\begin{table*}[!ht]
	\caption{Modelling results for the 8 positions selected from the scatter plots (App.~\ref{chisqplots}).}
	\begin{center}
	\begin{tabular}{c c c c c c c }
	\hline 
	\hline
		       &   H$_{2}$        &	Line   &	 Central  & non-LTE        & non-LTE   & LTE \\
	Position &  volume density  & width	 &  velocity 	& column density & T$_{\mathrm{kin}}$ 	& T$_{\mathrm{exc}}$\\
	 &($\times~10^{5}$~cm$^{-3}$)&(kms$^{-1}$)&(kms$^{-1}$)&($\times~10^{15}$~cm$^{-2}$)&(K)&(K)\\
	\hline
	NA & {2.60} & {1.6}	& {8.5} & {15.4}	& {11.7}		& {10.3}\\     
	NB & {2.25} & {1.5} & {8.4} & {11.9}	& {10.6}		& {9.5}\\ 			
	NC & {1.62} & {1.9} & {8.5} & {11.9}  & {10.5}		& {9.5}\\	 		
	ND & {6.50} & {1.4} & {8.5} & {13.5}	& {11.4}	  & {9.5}\\ 			
	\hline
	SA1 & {2.98} & {2.2} & {7.8} & {6.4}	& {18.2}		& {13.4}\\	    
	SA2 & {2.98} & {1.0} & {8.3} & {4.2}	& {6.6}		& \\		
	SB1 & {3.99} & {1.5} & {6.9} & {6.0}	& {11.9}		& {12.7}\\     
	SB2 & {3.99} & {1.1} & {8.5} & {4.4}	& {14.8}		& \\	
	SC1 & {6.28} & {1.7} & {7.7} & {5.8}	&	{11.7}		& {10.2}\\     
	SC2 &	{6.28} & {1.2} & {8.7} & {1.4}	& {14.2}	  & \\
	SD1 & {7.81} & {1.1} & {6.9} & {7.3}	&	{16.6}		& {12.8}\\     
	SD2 & {7.81} & {1.4} & {8.2} & {5.8}	& {14.4}		& \\
	\hline
	\end{tabular}
	\end{center}
	\begin{flushleft}
	{Positions starting with N are in the NW sub-cluster while positions
          in the southern region  start with S.  For the
          positions in the SE, the labels 1 and 2 identify the low and
          high velocity components respectively.}
  \end{flushleft}
	\label{tab:radex}
\end{table*}

The central velocity, line widths and integrated intensity of each transition
were retrieved from the data by fitting the average spectrum within a 5$''$
radius of each position. {The central velocity and line widths shown in
  Table~\ref{tab:radex} are the average over the three transitions, and have
  an uncertainty of the order of 0.1~kms$^{-1}$.}  For the 4 positions in the
NW sub-cluster, this procedure is straight-forward as the lines of all three
transitions are well represented by single Gaussians. However, the spectra of the SE
sub-cluster positions, having two velocity components, was
separately fitted with 2 Gaussians in order to
investigate any possible differences between the two components.  We used a
$\chi^{2}$ comparison to find the best fit of the RADEX models to the observed
{ratios $I_{(1-0)}$/$I_{(2-1)}$ and $I_{(1-0)}$/$I_{(3-2)}$ as well as the
  absolute value of $I_{(1-0)}$}.

In calculating the $\chi^{2}$, the relative errors of the input integrated
intensities were assumed to be equal to the relative errors in the intensity,
T$_{\mathrm{mb}}$), given by the r.m.s. of the line fit given by
\textsc{class}. The results from the RADEX models are presented in
Table~\ref{tab:radex}. For comparison, the table also includes the temperature
and column densities retrieved using the rotation diagram method. The $\chi^2$
surfaces are shown in App.~\ref{chisqplots}, while App.~\ref{intensities}
shows the input and best fit integrated intensities for the three lines, as
well as the implied abundances from the non-LTE analysis.

Table~\ref{tab:radex} shows that in the NW region the LTE (rotation diagram)
and non-LTE (RADEX) analyses are in good agreement. Both the rotation diagram
and RADEX show variations in T$_{\mathrm{ex}}$ of only 1~K. They produce absolute 
values of temperature which differ by at most {10\%}, and the trend
in temperature between positions is similar. 
Comparison of the C$^{18}$O
column densities and the H$_{2}$ column density calculated from the dust
continuum emission at the same positions, imply C$^{18}$O abundances a factor
of $\sim$2.5 smaller than typical values (App.~\ref{intensities}).  Given the
low optical depth of the C$^{18}$O emission (Sec.~\ref{scatter}), the low
C$^{18}$O column density, and hence abundance, implies that the C$^{18}$O is
depleted with respect to the molecular hydrogen, even in the warmer envelope
of SMM1, consistent with the results of \citet{1999ApJ...513..350H}.

Somewhat surprisingly given the low optical depth of the C$^{18}$O and despite
its depletion, neither the LTE nor the non-LTE analysis finds evidence of
increased temperatures towards the apparently warmer inner regions of the
embedded protostars. Although the dust emission indicates the presence of warm
dust towards the protostars \citep{1999MNRAS.309..141D}, the C$^{18}$O
emission implies low and uniform temperatures.  Although dilution of the warm
inner region within the 24$''$ beam may contribute to the difficulty in
detecting the warmest gas, it is surprising that no evidence of any
temperature increase is seen.  CO is predicted to freeze-out on to grain
surfaces at temperatures below $\sim18$K, consistent with the low C$^{18}$O
excitation temperature, but not at dust temperatures of, for instance,
$\sim30$K seen towards SMM1.  Since this is above the sublimation temperature
of pure CO ice it is possible the CO could be trapped in a water rich ice,
which would only sublime and return CO to the gas phase at temperatures of
$\sim$100~K \citep[e.g.][]{2009A&A...495..881V}.
  
The southern region is more complex. 
In terms of column density, for all four positions the non-LTE results show the LVC to have slightly higher 
column density that the HVC. We find that the LTE and non-LTE approaches agree
in the sense that the northern positions and SD have higher values of
C$^{18}$O total column density (summed over both velocity components where
necessary). These are followed with decreasing column density by SB, then SA and finally SC .  
  
SA is at the position of SMM4, where there are two components of the C$^{18}$O emission, a strong low velocity
component (SA1) plus a weaker high velocity component (SA2). In the
J=3$\rightarrow$2 transition, the high velocity component becomes faint and
difficult to separate from the lower velocity component (see
Table~\ref{tab:intens} in App.~\ref{intensities}).  This weak
J=3$\rightarrow$2 emission constrains the temperature to 6.6~K for the higher
velocity component (SA2). At SC the two cloud components are also
significantly blended. The temperature of the LVC at this position (SC1) is $\sim$11~K,
constrained within $\sim$1~K (App.~\ref{chisqplots}), with the weaker HVC warmer,
but somewhat less well constrained. 

In general, in the south, the lower velocity component has a higher temperature
toward the most central positions studied (SA and SD). Then at SB and SC, at
the edges of the dust emission, the temperatures are similar with both
components still higher than the temperatures generally found to the north
($\sim$12~K and 14~K in the south versus $\sim$11~K in the north).
Overall, and with the exception of SA, the HVC has higher temperatures 
than in the north, around $\sim$14~K. On the other hand, the LVC traces the 
temperature trend as identified by the LTE study (Fig.~\ref{fig:exc}) better than the HVC, but the
absolute LTE temperatures are between the non-LTE values for LVC and HVC.
 
Therefore, we conclude the temperature rise toward the south is real. Such a
rise is consistent with a scenario where this region is tracing the
interaction/collision between two clouds, with a shock layer with higher
temperatures and complex motions at the interface.

	
\section{Discussion}
	\label{discussion}	

\subsection{Two different sub-clusters in Serpens}

Our study of the Serpens Main Cluster has shown that two apparently very
similar protoclusters as seen in submillimetre dust continuum emission can
reveal very different dynamical and physical properties in molecular lines.
Despite all the outflows seen in $^{12}$CO in the region, the denser gas
around the cores seen in C$^{18}$O and C$^{17}$O does not seem to be perturbed
and is able to provide details of the quiescent material in the cloud.

In the NW sub-cluster the bulk of emission has a velocity around
8.5~kms$^{-1}$.  However, there is a lower velocity component of the gas east
of the sub-cluster (Fig.~\ref{fig:northslices}) with the transition between
these component being rather smooth. The velocity difference between the
submillimetre sources in this sub-cluster is small, ranging from 0.1 to
0.3~kms$^{-1}$.

The physical conditions in this NW sub-cluster are also rather coherent.
Temperatures and column densities derived from both LTE and non-LTE analyses
are consistent and show little variation within the sub-cluster.  The
C$^{18}$O emission peaks are mostly consistent with the dust
peaks. Clump-finding studies of this region retrieved two main peaks which are
directly related to the two stronger submillimetre sources in the NW
sub-cluster: SMM1 and SMM9. However there are no evident temperature peaks
associated with the submillimetre sources.  The remaining gas emission in the
sub-cluster is either associated with these main peaks or weaker structures
surrounding the main bulk of the dust emission. The gas column density very
closely follows the clumps/integrated intensity distribution of the gas,
particularly in the lower J transitions tracing the colder gas.

The SE sub-cluster on the other hand is a much richer region in its dynamics
and properties. There are two velocity components/clouds along the line of
sight, clearly identified using both clump-finding and position-velocity
diagrams. These two clouds appear to be interacting. They are more offset in
velocity in the south and start to mix moving to the north within the
sub-cluster. Furthermore, the submillimetre sources in this region appear at
the edges of the interface of the two components, whereas the dust filament
appears in the interface. 
Most of the southern
{submillimetre} sources appear to have a stronger association with the HVC,
despite having some emission from the LVC along the same line of sight. A
counterexample however is SMM2 which, as can be seen in the PV diagrams, has a
stronger C$^{18}$O J=1$\rightarrow$0 lower velocity component. The overall
dust filament, as seen in 850~$\mu$m, coincides with the N-S lane where the
two components overlap, suggesting it is tracing the interface region between
the components, the region where they are interacting. Ultimately, this
interaction might have been responsible for triggering the star formation
episode in the SE sub-cluster.
 
In contrast to the NW sub-cluster, the LTE temperature in the SE sub-cluster
is both higher and more structured, peaking close to the ridge of dust
continuum emission.  Unlike the north, the two velocity components in the
south are difficult to fit with a single well defined temperature.  The
general trend, however, points to higher temperatures in the southern
sub-cluster than in the northern sub-cluster.

The modelled column density map (Fig.~\ref{fig:columndens}) closely traces the
emission from the lower transitions (J=1$\rightarrow$0). The high C$^{18}$O
column density regions in the SE are not associated with any of the submillimetre
sources, but rather the southern filament. The region
with enhanced temperature, however, does not coincide with the highest column
density regions. The uniform dust emission over this SE region results from
the southern filament having lower temperature but higher column density
whereas the northern part of the SE sub-cluster is slightly less dense, but
warmer, resulting in equivalent 850~$\mu$m dust emission.

\subsection{A Proposed Scenario}

The velocity, temperature and density structure of Serpens suggest a more
complex picture than simple rotation which has previously been invoked to
explain the velocity structure \citep[e.g.][]{2002A&A...392.1053O}. 

It is known that cloud-cloud or flow collisions happen in the Galaxy as
molecular clouds move within the spiral arms. Furthermore, simulations of
cloud-cloud collisions \citep[e.g.][]{2007MNRAS.378..507K} have shown that
density enhancements in the collision layers can be high enough to trigger
star formation. Additionally, clouds are commonly seen as filamentary
structures, not only during, but also prior, to star formation. We suggest
that the two velocity components seen in Serpens are tracing two clouds along
the line of sight and that the interaction of these clouds is a key ingredient
in the star formation in Serpens.

We propose that we are seeing two somewhat filamentary clouds traveling toward
each other and colliding where the southern sub-cluster is being formed. The
cloud coming toward us is to the east while the cloud moving away from us is
to the west, and represents the main cloud. An inclination angle between the
two filaments could explain why the two velocity components are spatially
offset in the north but overlapping in the south.  This scenario explains both
the double peaked profiles of the optically thin lines and their distribution
along what has previously been identified as the `rotation axis' of this
region.

If the north region was initially close to collapse, the direct collision of
the clouds in the south could indirectly trigger or speed up this collapse in
the north without significantly enhancing the temperature or perturbing the
intrinsic, `well behaved' velocity and column density structure.  In the
south, however, such a collision makes it easy to understand why the density
and temperature enhancements are not necessarily associated with the sources,
as they are being generated by an external trigger: the collision.

Note in addition, that in the south, unlike the majority of the sources in the
north, there is a poor correlation between the submillimetre sources
\citep{1999MNRAS.309..141D} and 24~$\mu$m sources \citep{2007ApJ...663.1139H},
as shown in Fig.~\ref{fig:srp}, suggesting a wider spread of ages {of the
  protostars in the south} than in the north. Such an age spread would be
consistent with a collision in the sense that a collision is not a one-off
event but rather an ongoing process.



A first test to this collision scenario is the timescale for which such clouds
would cross each other, their interaction time. We assume that each cloud is a
filament of radius of 0.1~pc (similar to the size of the dust 850~$\mu$m
emission). In addition we adopt a collision velocity of 1~km\ s$^{-1}$
(approximately the mean observed velocity difference, along the line of sight,
between the two components). The timescale from when the clouds start
colliding until they are completely separated, assuming a head on collision,
is $4 \times 10^{5}$~years, consistent with the estimated $\sim10^5$~year age
of the region \citep{2007ApJ...663.1149H,2004A&A...421..623K}.


Several simulations of cloud collisions such as proposed here exist in the
literature. For example, SPH simulations of clump-clump collisions from
\citet{2007MNRAS.378..507K}, have shown that two approaching clumps with a
slow collision velocity of 1~kms$^{-1}$ (Mach number of 5), can indeed trigger
star formation in the collision layer. 
Specific simulations of the proposed collision in Serpens will be presented in
a subsequent paper (Duarte-Cabral et al. 2010 in prep.)

\section{Conclusions}
\label{conc}

The dynamics and structure of the Serpens Main Cluster have been studied in
detail, as an example of a complex low mass star cluster forming region. This
study has provided a view of the dynamics and structure of the region.  The
Main Cluster comprises a very young star forming region subdivided into two
sub-clusters.  A careful investigation of the clump structure and excitation
in the region shows that these two sub-clusters have similar overall masses
but quite different properties.

The NW sub-cluster is homogeneous in velocity and temperature structure, with
the submillimetre sources well correlated with the gas peaks as well as with
the 24~$\mu$m sources. On the other hand, in the SE sub-cluster there are two
velocity components in the gas and the gas temperature is more variable.  The
gas column density and temperature peaks do not coincide with the
submillimetre sources, but rather lay in the regions between
them. Furthermore, the 24~$\mu$m sources in the south are poorly correlated
with the dust emission.

Our analysis suggests a scenario of cloud-cloud collision triggering the star
formation in the SE cub-cluster, potentially inducing perturbations which
indirectly affect the NW sub-cluster, hastening somewhat its collapse. SPH
simulations of this scenario, to better understand how well it can reproduce
observables, such as the velocity profile and the column density properties of
the region, will be presented in a future paper.

\begin{verbatim}


\end{verbatim}

We thank the anonymous referee for helpful comments which helped improve 
this paper. Ana Duarte Cabral is funded by the Funda{\c{c}}{\~a}o para a 
Ci{\^e}ncia e a Tecnologia of Portugal, under the grant reference SFRH/BD/36692/2007. 
IRAM is supported by INSU/CNRS (France), MPG (Germany), and IGN (Spain). The data
reduction and analysis was done using the GILDAS software
(http://www.iram.fr/IRAMFR/GILDAS) and the Starlink software
(http://starlink.jach.hawaii.edu/starlink). The James Clerk Maxwell Telescope
(JCMT) is operated by the Joint Astronomy Centre, on behalf of the Particle
Physics and Astronomy Research Council of the United Kingdom, the Netherlands
Organization for Scientific Research, and the National Research Council of
Canada. This research used the facilities of the Canadian Astronomy Data
Centre operated by the National Research Council of Canada with the support of
the Canadian Space Agency.


\bibliographystyle{aa}	
\bibliography{ref}		

\begin{thebibliography}{38}
\expandafter\ifx\csname natexlab\endcsname\relax\def\natexlab#1{#1}\fi

\bibitem[{{Blitz}(1993)}]{1993prpl.conf..125B}
{Blitz}, L. 1993, in Protostars and Planets III, ed. {E.~H.~Levy \&
  J.~I.~Lunine}, 125--161

\bibitem[{{Cambr{\'e}sy}(1999)}]{1999A&A...345..965C}
{Cambr{\'e}sy}, L. 1999, \aap, 345, 965

\bibitem[{{Curtis} {et~al.}(2009){Curtis}, {Richer}, \&
  {Buckle}}]{2009MNRAS.tmp.1566C}
{Curtis}, E.~I., {Richer}, J.~S., \& {Buckle}, J.~V. 2009, \mnras, 1566

\bibitem[{{Davis} {et~al.}(2000){Davis}, {Chrysostomou}, {Matthews}, {Jenness},
  \& {Ray}}]{2000ApJ...530L.115D}
{Davis}, C.~J., {Chrysostomou}, A., {Matthews}, H.~E., {Jenness}, T., \& {Ray},
  T.~P. 2000, \apjl, 530, L115

\bibitem[{{Davis} {et~al.}(1999){Davis}, {Matthews}, {Ray}, {Dent}, \&
  {Richer}}]{1999MNRAS.309..141D}
{Davis}, C.~J., {Matthews}, H.~E., {Ray}, T.~P., {Dent}, W.~R.~F., \& {Richer},
  J.~S. 1999, \mnras, 309, 141

\bibitem[{{Di Francesco} {et~al.}(2008){Di Francesco}, {Johnstone}, {Kirk},
  {MacKenzie}, \& {Ledwosinska}}]{2008ApJS..175..277D}
{Di Francesco}, J., {Johnstone}, D., {Kirk}, H., {MacKenzie}, T., \&
  {Ledwosinska}, E. 2008, \apjs, 175, 277

\bibitem[{{Eiroa} {et~al.}(1992){Eiroa}, {Torrelles}, {Gomez}, {Sakamoto},
  {Hasegawa}, {Kawabe}, {Hayashi}, \& {Casali}}]{1992PASJ...44..155E}
{Eiroa}, C., {Torrelles}, J.~M., {Gomez}, J.~F., {et~al.} 1992, \pasj, 44, 155

\bibitem[{{Enoch} {et~al.}(2007){Enoch}, {Glenn}, {Evans}, {Sargent}, {Young},
  \& {Huard}}]{2007ApJ...666..982E}
{Enoch}, M.~L., {Glenn}, J., {Evans}, II, N.~J., {et~al.} 2007, \apj, 666, 982

\bibitem[{{Frerking} {et~al.}(1982){Frerking}, {Langer}, \&
  {Wilson}}]{1982ApJ...262..590F}
{Frerking}, M.~A., {Langer}, W.~D., \& {Wilson}, R.~W. 1982, \apj, 262, 590

\bibitem[{{Fuller} \& {Ladd}(2002)}]{2002ApJ...573..699F}
{Fuller}, G.~A. \& {Ladd}, E.~F. 2002, \apj, 573, 699

\bibitem[{{Fuller} \& {Myers}(1993)}]{1993ApJ...418..273F}
{Fuller}, G.~A. \& {Myers}, P.~C. 1993, \apj, 418, 273

\bibitem[{{Graves} {et~al.}(2010){Graves}, {Richer}, {Buckle}, {Duarte-Cabral},
  {Fuller}, {Brunt}, {Butner}, {Cavanagh}, {Chrysostomou}, {Davis}, {Etxaluze},
  {di Francesco}, {Friberg}, {Friesen}, {Greaves}, {Hogerheijde}, {Johnstone},
  {Matthews}, {Matthews}, {Nutter}, {Rawlings}, {Sadavoy}, {Simpson},
  {Tothill}, {Tsamis}, {Viti}, {Ward-Thompson}, {Wouterloot}, \&
  {Yates}}]{2010MNRAS...Graves}
{Graves}, S.~F., {Richer}, J.~S., {Buckle}, J.~V., {et~al.} 2010, \mnras,
  in press

\bibitem[{{Harvey} {et~al.}(2007{\natexlab{a}}){Harvey}, {Mer{\'{\i}}n},
  {Huard}, {Rebull}, {Chapman}, {Evans}, \& {Myers}}]{2007ApJ...663.1149H}
{Harvey}, P., {Mer{\'{\i}}n}, B., {Huard}, T.~L., {et~al.} 2007{\natexlab{a}},
  \apj, 663, 1149

\bibitem[{{Harvey} {et~al.}(2007{\natexlab{b}}){Harvey}, {Rebull}, {Brooke},
  {Spiesman}, {Chapman}, {Huard}, {Evans}, {Cieza}, {Lai}, {Allen}, {Mundy},
  {Padgett}, {Sargent}, {Stapelfeldt}, {Myers}, {van Dishoeck}, {Blake}, \&
  {Koerner}}]{2007ApJ...663.1139H}
{Harvey}, P.~M., {Rebull}, L.~M., {Brooke}, T., {et~al.} 2007{\natexlab{b}},
  \apj, 663, 1139

\bibitem[{{Hodapp}(1999)}]{1999AJ....118.1338H}
{Hodapp}, K.~W. 1999, \aj, 118, 1338

\bibitem[{{Hogerheijde} {et~al.}(1999){Hogerheijde}, {van Dishoeck},
  {Salverda}, \& {Blake}}]{1999ApJ...513..350H}
{Hogerheijde}, M.~R., {van Dishoeck}, E.~F., {Salverda}, J.~M., \& {Blake},
  G.~A. 1999, \apj, 513, 350

\bibitem[{{Hurt} \& {Barsony}(1996)}]{1996ApJ...460L..45H}
{Hurt}, R.~L. \& {Barsony}, M. 1996, \apjl, 460, L45+

\bibitem[{{Hurt} {et~al.}(1996){Hurt}, {Barsony}, \&
  {Wootten}}]{1996ApJ...456..686H}
{Hurt}, R.~L., {Barsony}, M., \& {Wootten}, A. 1996, \apj, 456, 686

\bibitem[{{Johnstone} \& {Bally}(2006)}]{2006ApJ...653..383J}
{Johnstone}, D. \& {Bally}, J. 2006, \apj, 653, 383

\bibitem[{{J{\o}rgensen} {et~al.}(2002){J{\o}rgensen}, {Sch{\"o}ier}, \& {van
  Dishoeck}}]{2002A&A...389..908J}
{J{\o}rgensen}, J.~K., {Sch{\"o}ier}, F.~L., \& {van Dishoeck}, E.~F. 2002,
  \aap, 389, 908

\bibitem[{{Kaas} {et~al.}(2004){Kaas}, {Olofsson}, {Bontemps}, {Andr{\'e}},
  {Nordh}, {Huldtgren}, {Prusti}, {Persi}, {Delgado}, {Motte}, {Abergel},
  {Boulanger}, {Burgdorf}, {Casali}, {Cesarsky}, {Davies}, {Falgarone},
  {Montmerle}, {Perault}, {Puget}, \& {Sibille}}]{2004A&A...421..623K}
{Kaas}, A.~A., {Olofsson}, G., {Bontemps}, S., {et~al.} 2004, Astronomy and
  Astrophysics, 421, 623

\bibitem[{{Kitsionas} \& {Whitworth}(2007)}]{2007MNRAS.378..507K}
{Kitsionas}, S. \& {Whitworth}, A.~P. 2007, \mnras, 378, 507

\bibitem[{{Lada} \& {Lada}(2003)}]{2003ARA&A..41...57L}
{Lada}, C.~J. \& {Lada}, E.~A. 2003, \araa, 41, 57

\bibitem[{{Ladd} {et~al.}(1998){Ladd}, {Fuller}, \&
  {Deane}}]{1998ApJ...495..871L}
{Ladd}, E.~F., {Fuller}, G.~A., \& {Deane}, J.~R. 1998, \apj, 495, 871

\bibitem[{{Olmi} \& {Testi}(2002)}]{2002A&A...392.1053O}
{Olmi}, L. \& {Testi}, L. 2002, \aap, 392, 1053

\bibitem[{{Penzias}(1980)}]{1980IAUS...87..397P}
{Penzias}, A.~A. 1980, in IAU Symposium, Vol.~87, Interstellar Molecules, ed.
  {B.~H.~Andrew}, 397--402

\bibitem[{{Peretto} {et~al.}(2006){Peretto}, {Andr{\'e}}, \&
  {Belloche}}]{2006A&A...445..979P}
{Peretto}, N., {Andr{\'e}}, P., \& {Belloche}, A. 2006, \aap, 445, 979

\bibitem[{{Pineda} {et~al.}(2009){Pineda}, {Rosolowsky}, \&
  {Goodman}}]{2009ApJ...699L.134P}
{Pineda}, J.~E., {Rosolowsky}, E.~W., \& {Goodman}, A.~A. 2009, \apjl, 699,
  L134

\bibitem[{{Schoeier} {et~al.}(2005){Schoeier}, {van der Tak}, {van Dishoeck},
  \& {Black}}]{2005yCat..34320369S}
{Schoeier}, F.~L., {van der Tak}, F.~F.~S., {van Dishoeck}, E.~F., \& {Black},
  J.~H. 2005, VizieR Online Data Catalog, 343, 20369

\bibitem[{{Strai{\v z}ys} {et~al.}(1996){Strai{\v z}ys}, {{\v C}ernis}, \&
  {Barta{\v s}i{\= u}te}}]{1996BaltA...5..125S}
{Strai{\v z}ys}, V., {{\v C}ernis}, K., \& {Barta{\v s}i{\= u}te}, S. 1996,
  Baltic Astronomy, 5, 125

\bibitem[{{van der Tak} {et~al.}(2007){van der Tak}, {Black}, {Sch{\"o}ier},
  {Jansen}, \& {van Dishoeck}}]{2007A&A...468..627V}
{van der Tak}, F.~F.~S., {Black}, J.~H., {Sch{\"o}ier}, F.~L., {Jansen}, D.~J.,
  \& {van Dishoeck}, E.~F. 2007, \aap, 468, 627

\bibitem[{{van der Tak} {et~al.}(1999){van der Tak}, {van Dishoeck}, {Evans},
  {Bakker}, \& {Blake}}]{1999ApJ...522..991V}
{van der Tak}, F.~F.~S., {van Dishoeck}, E.~F., {Evans}, II, N.~J., {Bakker},
  E.~J., \& {Blake}, G.~A. 1999, \apj, 522, 991

\bibitem[{{Visser} {et~al.}(2009){Visser}, {van Dishoeck}, {Doty}, \&
  {Dullemond}}]{2009A&A...495..881V}
{Visser}, R., {van Dishoeck}, E.~F., {Doty}, S.~D., \& {Dullemond}, C.~P. 2009,
  \aap, 495, 881

\bibitem[{{Walsh} {et~al.}(2007){Walsh}, {Myers}, {Di Francesco}, {Mohanty},
  {Bourke}, {Gutermuth}, \& {Wilner}}]{2007ApJ...655..958W}
{Walsh}, A.~J., {Myers}, P.~C., {Di Francesco}, J., {et~al.} 2007, \apj, 655,
  958

\bibitem[{{Ward-Thompson} {et~al.}(2007){Ward-Thompson}, {Di Francesco},
  {Hatchell}, {Hogerheijde}, {Nutter}, {Bastien}, {Basu}, {Bonnell}, {Bowey},
  {Brunt}, {Buckle}, {Butner}, {Cavanagh}, {Chrysostomou}, {Curtis}, {Davis},
  {Dent}, {van Dishoeck}, {Edmunds}, {Fich}, {Fiege}, {Fissel}, {Friberg},
  {Friesen}, {Frieswijk}, {Fuller}, {Gosling}, {Graves}, {Greaves}, {Helmich},
  {Hills}, {Holland}, {Houde}, {Jayawardhana}, {Johnstone}, {Joncas}, {Kirk},
  {Kirk}, {Knee}, {Matthews}, {Matthews}, {Matzner}, {Moriarty-Schieven},
  {Naylor}, {Padman}, {Plume}, {Rawlings}, {Redman}, {Reid}, {Richer},
  {Shipman}, {Simpson}, {Spaans}, {Stamatellos}, {Tsamis}, {Viti}, {Weferling},
  {White}, {Whitworth}, {Wouterloot}, {Yates}, \& {Zhu}}]{2007PASP..119..855W}
{Ward-Thompson}, D., {Di Francesco}, J., {Hatchell}, J., {et~al.} 2007, \pasp,
  119, 855

\bibitem[{{Williams} {et~al.}(1994){Williams}, {de Geus}, \&
  {Blitz}}]{1994ApJ...428..693W}
{Williams}, J.~P., {de Geus}, E.~J., \& {Blitz}, L. 1994, \apj, 428, 693

\bibitem[{{Williams} \& {Myers}(1999)}]{1999ApJ...518L..37W}
{Williams}, J.~P. \& {Myers}, P.~C. 1999, \apjl, 518, L37

\bibitem[{{Williams} \& {Myers}(2000)}]{2000ApJ...537..891W}
{Williams}, J.~P. \& {Myers}, P.~C. 2000, \apj, 537, 891

\end{thebibliography}

\onecolumn
\newpage

\begin{appendix}

\section{$\chi^{2}$ surfaces for RADEX fits}
\label{chisqplots}

\begin{figure}[!htb]
\centering
    \includegraphics[width=0.37\textwidth]{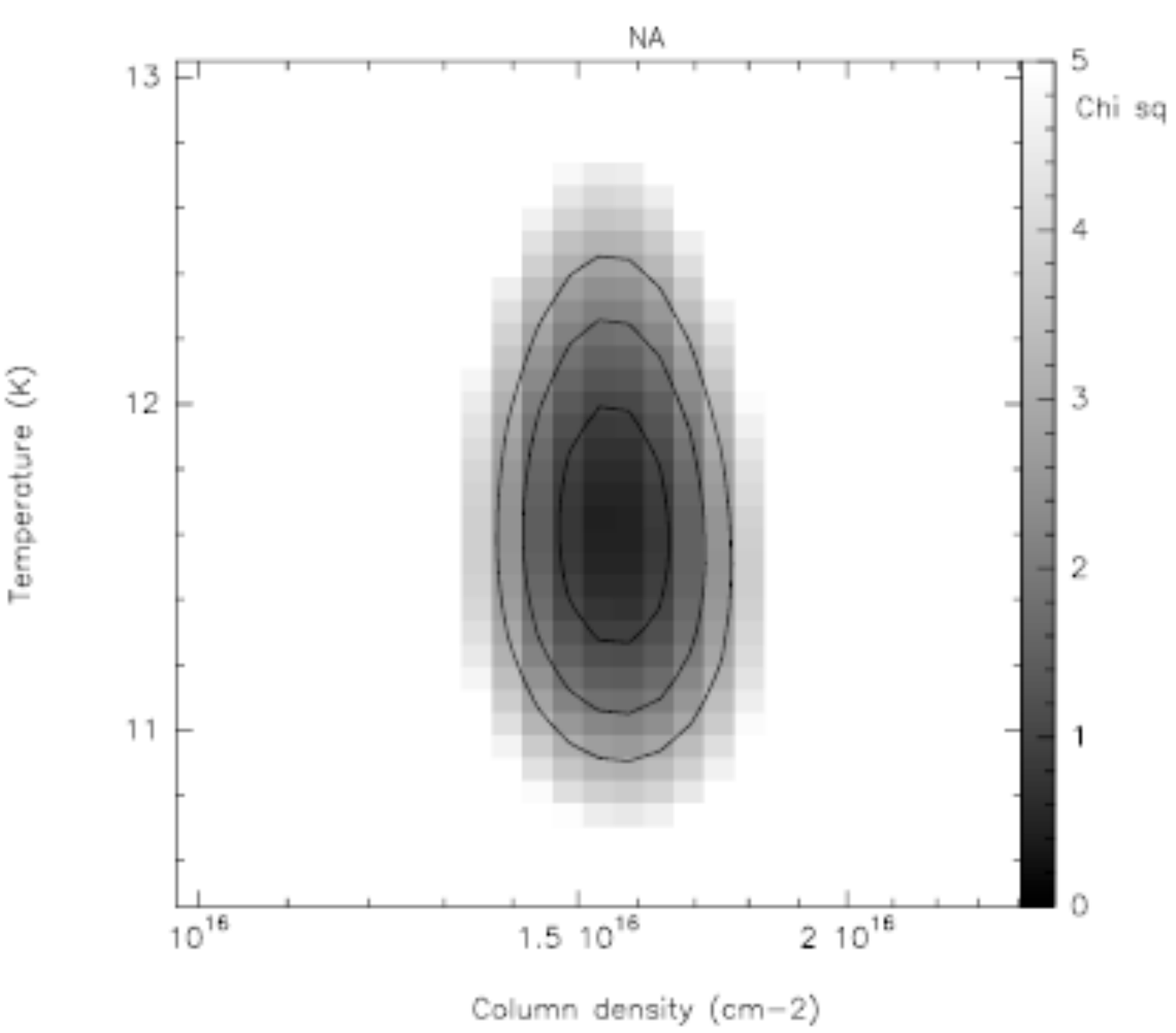}
    \caption{$\chi^{2}$ surface for the integrated intensity ratios at position
      NA.}
    \label{fig:NA}
		\includegraphics[width=0.37\textwidth]{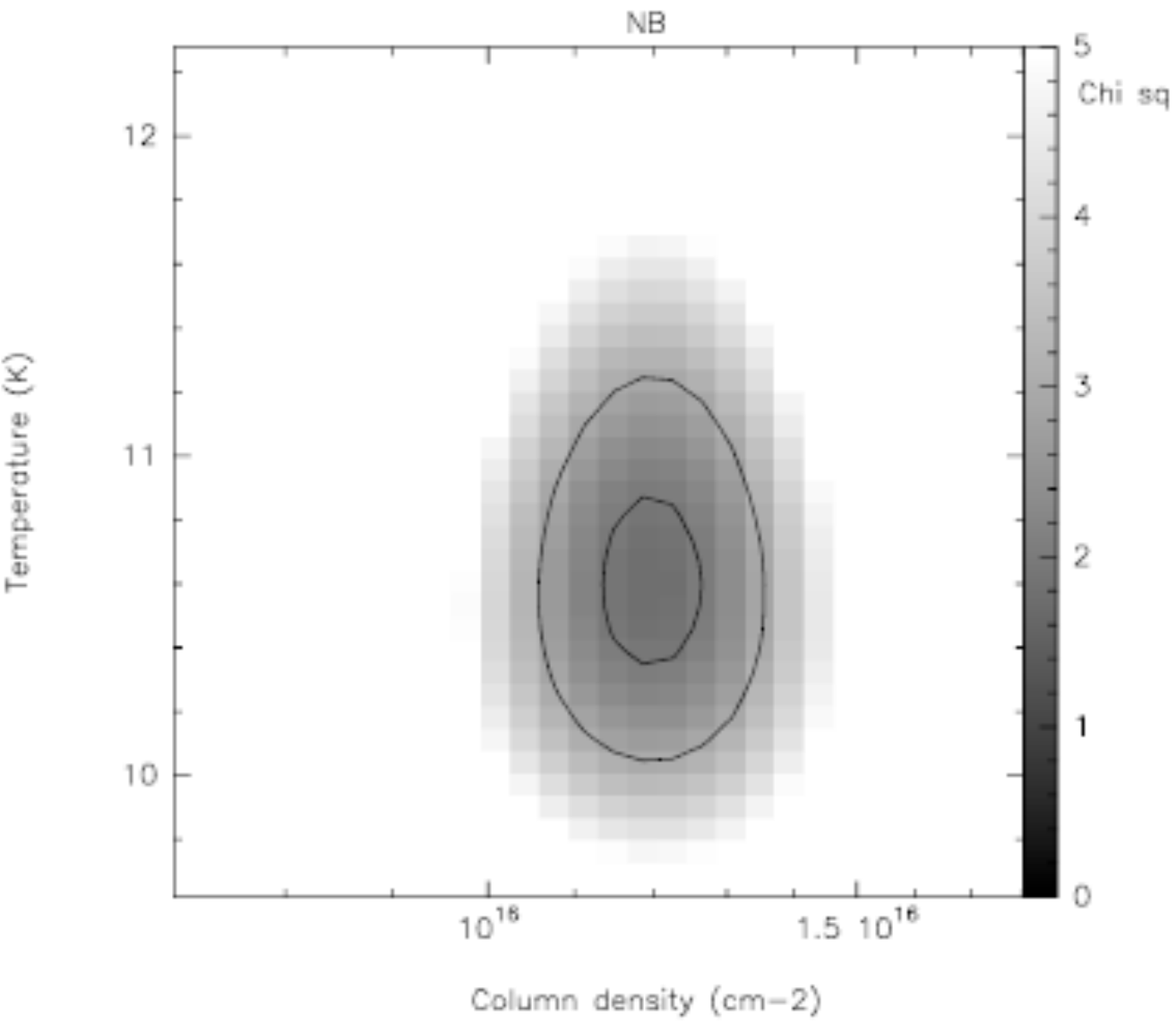}
		\caption{$\chi^{2}$ surface for the integrated intensity
                  ratios at position NB.}
		\label{fig:NB}
\end{figure}

\begin{figure}[!htb]
\centering
    \includegraphics[width=0.37\textwidth]{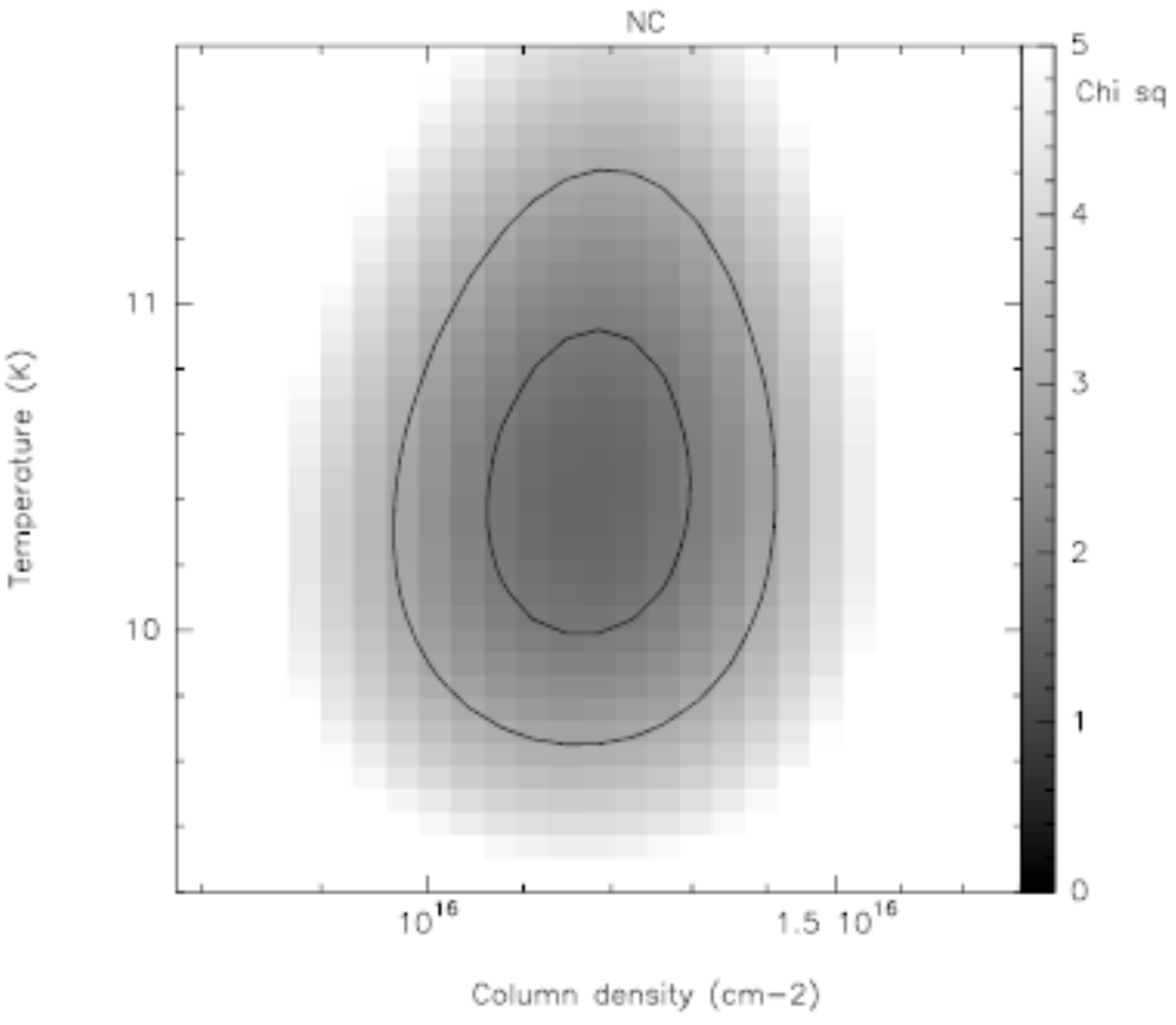}
    \caption{$\chi^{2}$ surface for the integrated intensity ratios at
      position NC.}
    \label{fig:NC}
    \begin{verbatim}
    
    \end{verbatim}
		\includegraphics[width=0.37\textwidth]{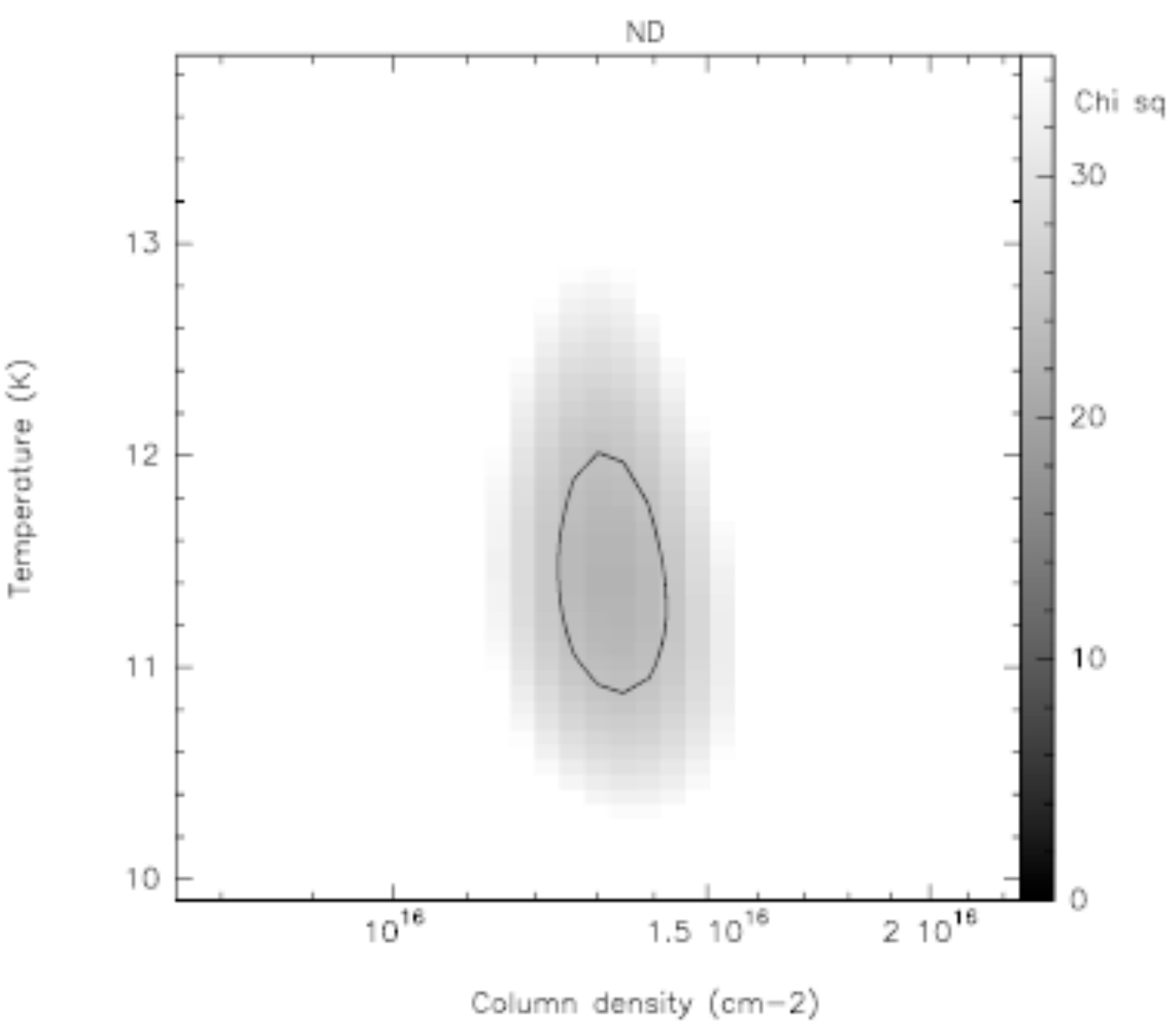}
		\caption{$\chi^{2}$ surface for the integrated intensity
                  ratios at position ND. Contour at 25.}
		\label{fig:ND}
\end{figure}

\begin{figure*}[!htb]
		\centering
    \includegraphics[width=0.8\textwidth]{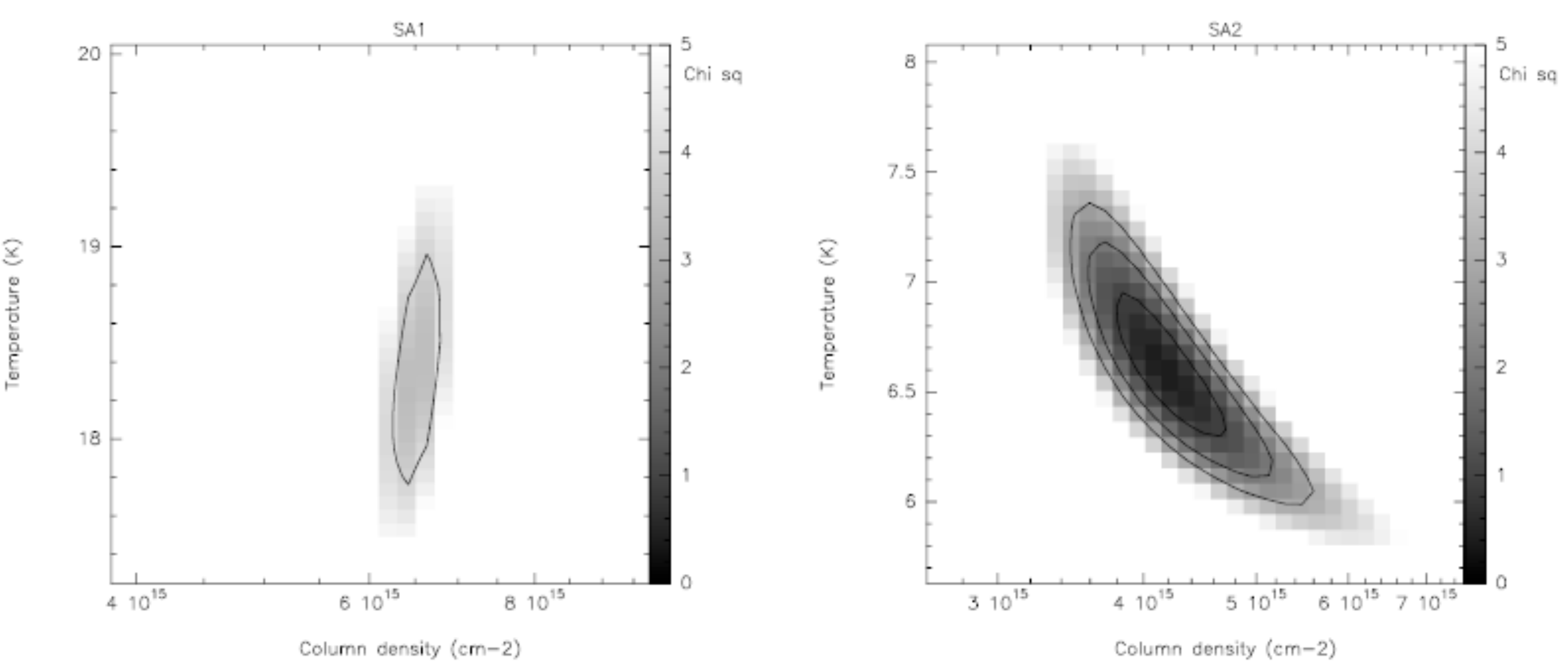}
    \hfill
    \caption{$\chi^{2}$ surfaces for the integrated intensity ratios at
      position SA: SA1 (LVC) on the left, and SA2 (HVC) on the right.}
    \label{fig:SA}
\end{figure*}

\begin{figure*}[!htb]
		\centering
    \includegraphics[width=0.8\textwidth]{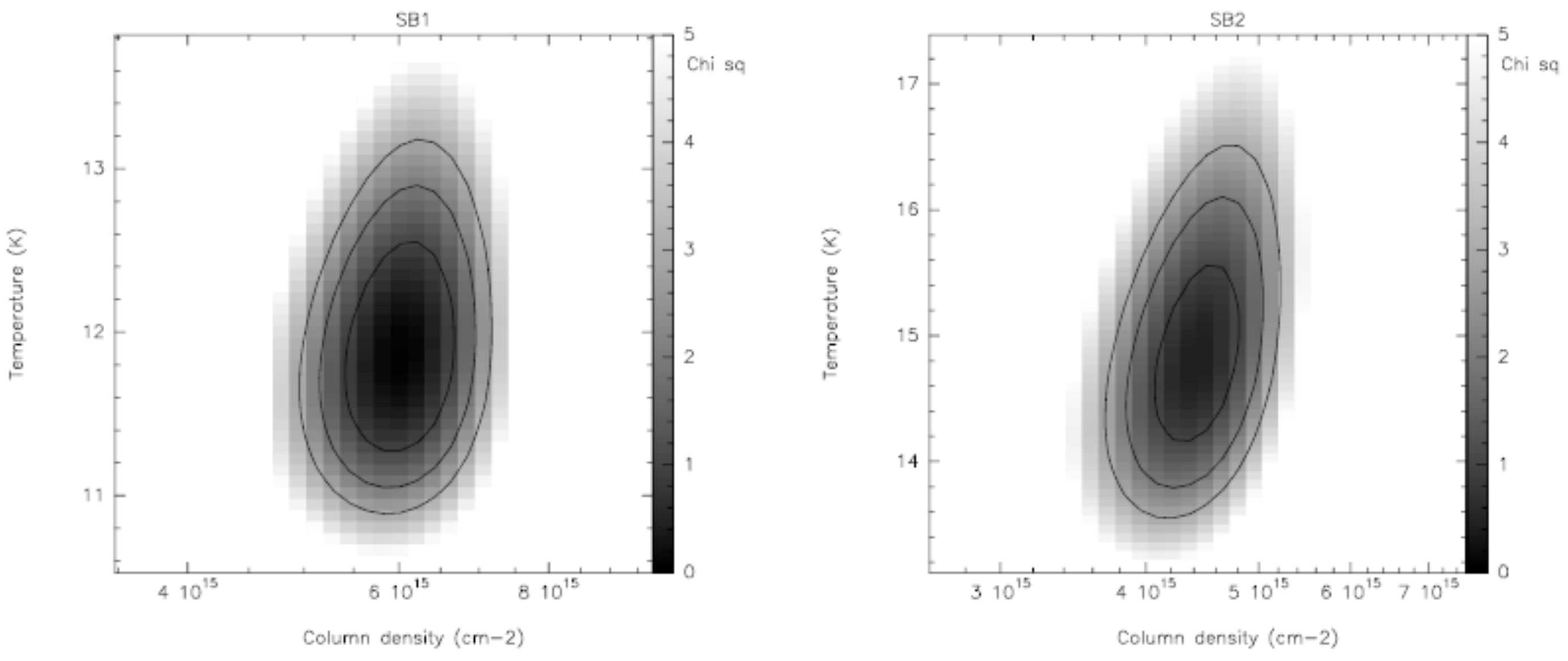}
    \hfill
    \caption{$\chi^{2}$ surface for the integrated intensity ratios at
      position SB: SB1
      (LVC) on the left, and SB2 (HVC) on the right.}
    \label{fig:SB}
\end{figure*}

\begin{figure*}[!htb] 
		\centering
    \includegraphics[width=.8\textwidth]{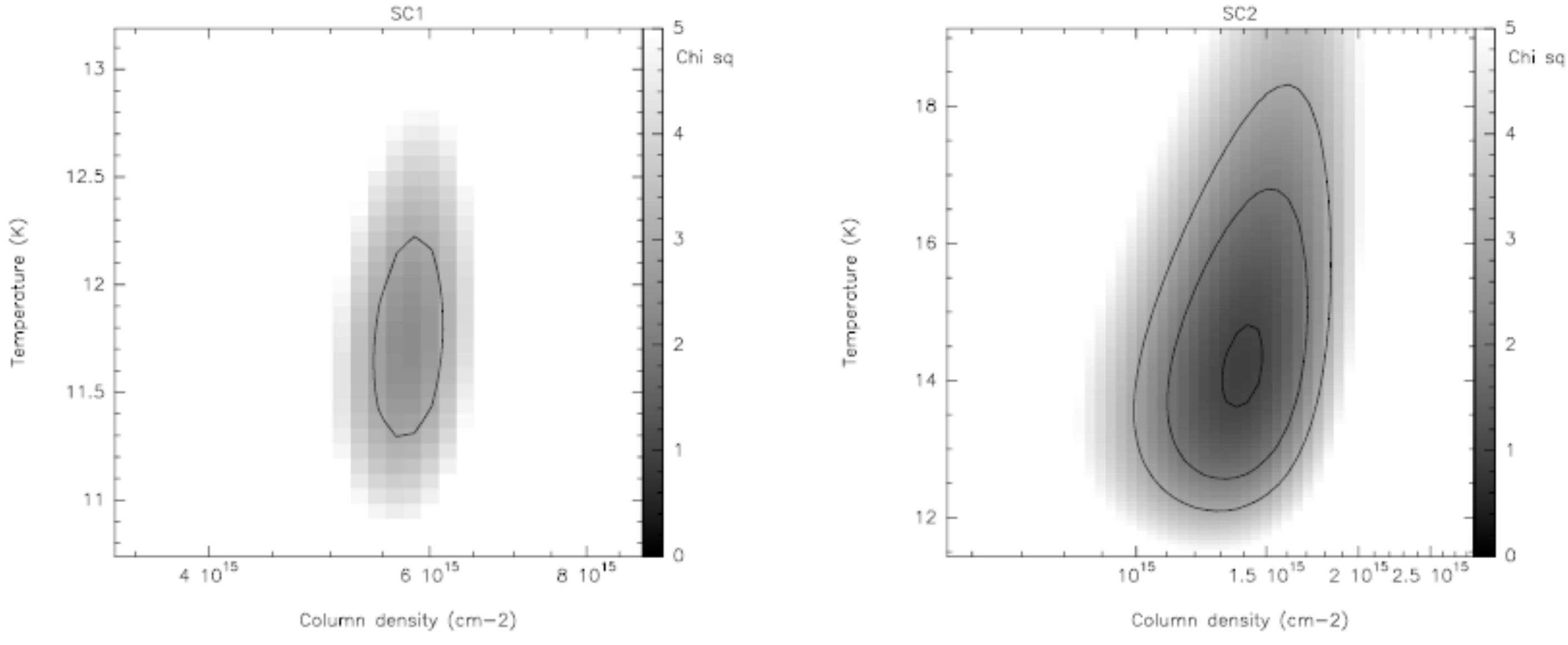}
    \hfill
    \caption{$\chi^{2}$ surface for the integrated intensity ratios at
      position SC: SC1
      (LVC) on the left and SC2 (HVC) on the right.}
    \label{fig:SC}      
\end{figure*}

\begin{figure*}[!htb]   
		\centering
    \includegraphics[width=.8\textwidth]{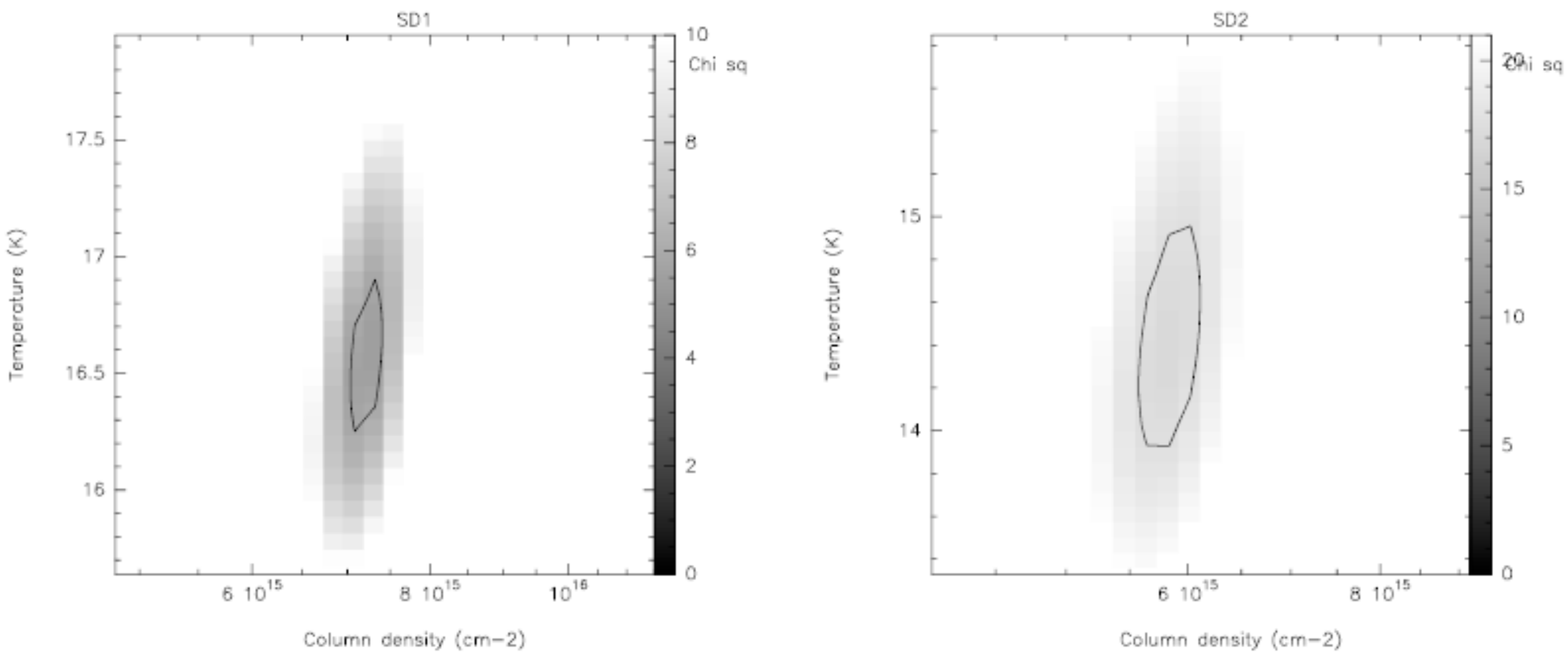}
    \hfill
    \caption{$\chi^{2}$ surface for the integrated intensity ratio at position
      SD: SD1 (LVC) on the left, and SD2 (HVC) on the right. Note the
      different colour scale for SC2.  Contours are 6 for SD1 and 18 for SD2}
    \label{fig:SD}
\end{figure*}

Figures~\ref{fig:NA} to \ref{fig:SD}) show the $\chi^{2}$ surfaces from the
non-LTE analysis of the line integrated intensity ratios for each of the
positions studied.  The $\chi^{2}$ has been calculated using
Eq.~(\ref{eq:chisq}), where $I_{(J_{up}-J_{low})}^{\mathrm{obs}}$ is the
observed integrated intensity of the transition between J$_{up}$ and
J$_{low}$, $\Delta$(x) is the uncertainty on the quantity $x$ and, finally,
$I_{(J_{up}-J_{low})}^{\mathrm{radex}}$ is the integrated intensity for each
transition as modelled by RADEX, for each combination of temperature and gas
column density.

\begin{equation}
\chi^{2} = \left(\frac{I_{(1-0)}^{\mathrm{obs}}/I_{(2-1)}^{\mathrm{obs}} - I_{(1-0)}^{\mathrm{radex}}/I_{(2-1)}^{\mathrm{radex}}}{\Delta(I_{(1-0)}^{\mathrm{obs}}/I_{(2-1)}^{\mathrm{obs}})}\right)^{2} + \left(\frac{I_{(1-0)}^{\mathrm{obs}}/I_{(3-2)}^{\mathrm{obs}} - I_{(1-0)}^{\mathrm{radex}}/I_{(3-2)}^{\mathrm{radex}}}{\Delta(I_{(1-0)}^{\mathrm{obs}}/I_{(3-2)}^{\mathrm{obs}})}\right)^{2} +
\left(\frac{I_{(1-0)}^{\mathrm{obs}} - I_{(1-0)}^{\mathrm{radex}}}{\Delta(I_{(1-0)}^{\mathrm{obs}})}\right)^{2}
\label{eq:chisq}
\end{equation}

For each case, the $\chi^{2}$ is plotted as a function of the RADEX output
temperature and gas column density. Given the use of 3 quantities in the fit,
we consider $\chi^{2}<3$, i.e. the reduced-$\chi^{2}<1$ a good fit. All
figures have contours at $\chi^{2}=1$, 2 and 3 with the exception of ND
(Fig.~\ref{fig:ND}) and SD (Fig.~\ref{fig:SD}).

\begin{verbatim}

\end{verbatim}

\vfill

\newpage

\begin{verbatim}

\end{verbatim}

\newpage

\begin{verbatim}

\end{verbatim}

\begin{verbatim}

\end{verbatim}

\newpage

\begin{verbatim}

\end{verbatim}

\begin{verbatim}

\end{verbatim}

\newpage

\section{C$^{18}$O intensities and abundances}
\label{intensities}

Table~\ref{tab:intens} presents the best fit integrated intensities from the
non-LTE (RADEX) modelling (\S~\ref{chisqplots}) , together with the observed
values, and the implied C$^{18}$O abundances.

\begin{table*}[!ht]
	\caption{Modelled integrated intensities and resulting abundances.}
	\begin{center}
	\footnotesize
	\begin{tabular}{c | c c c | c c c | c c c}
	\hline 
	\hline
		       &						  &	 Observed   &             &						  &	 Best fit   &        &   H$_{2}$    & C$^{18}$O  & C$^{18}$O \\
	Position & I$_{(1-0)}$ & I$_{(2-1)}$	& I$_{(3-2)}$ & I$_{(1-0)}$ & I$_{(2-1)}$	& I$_{(3-2)}$& column  &  column & fractional \\
	 & 					&(K~kms$^{-1}$) & 	  	 & 					&(K~kms$^{-1}$) &     &  density & density &  abundance\\
	 	       &				&				&			 &			&				&				&		($10^{23}$cm$^{-2}$)	 &	($10^{15}$cm$^{-2}$)	&	($\times10^{-8}$)\\
	\hline
	NA  & 9.15	& 10.65	& 6.94 & 9.23 & 10.42 & 7.28 	& 1.60 & 15.4	 & 9.6 \\    
	NB  &	7.47	& 8.87	& 4.76 & 7.44 & 8.15	& 5.20	& 1.39 & 11.9	 & 8.6 \\     
	NC  & 8.14	& 10.35	& 4.90 & 8.13	& 9.29	& 5.57	& 1.00 & 11.9	 & 12.0\\	    
	ND  & 7.90	& 10.09	& 4.82 & 7.99 & 8.88	& 6.20  & 4.01 & 13.5	 & 3.4 \\ 		
	\hline
	SA1  & 4.69	& 10.57	& 8.46 & 4.75 & 9.84	& 9.01	& 1.84 & 6.4	  & 5.8\\	     
	SA2  & 2.42	& 1.98	&	0.50 & 2.50 & 2.02	& 0.64	& 		 & 4.2		& 	 \\		
	SB1  & 4.63	& 6.66	& 4.20 & 4.74 & 6.72  & 4.37 	&	2.46 & 6.0	  & 4.2\\       
	SB2  & 3.32	& 5.99	& 4.31 & 3.35 & 5.73  & 4.50 	&      & 4.4		&		 \\	
	SC1  & 4.61	& 7.37	& 3.87 & 4.72 & 6.82  & 4.37  & 3.87 & 5.8		&	1.9\\       
	SC2  & 1.17	& 2.58	& 1.45 & 1.21 & 2.23  & 1.71 	&	     & 1.4		& 	 \\
	SD1  & 4.81	& 7.44	& 7.31 & 5.01 & 8.53  & 7.34 	& 4.82 & 7.3		&	2.7\\       
	SD2  & 4.41	& 8.66	& 4.66 & 4.45 & 7.42  & 5.74  &    	 & 5.8		& 	 \\
	\hline
	\end{tabular}
	\end{center}
	\begin{flushleft}
	{\footnotesize{Positions of the NW sub-cluster are identified as
            starting with N, while the south positions start with S.  For the
            positions in the SE, the labels 1 and 2 identify the lower (LVC) and
            higher velocity (HVC) components respectively. Where
            there are two velocity component lines,  the C$^{18}$O fractional 
            abundance was calculated using 
            the total column density of 
            C$^{18}$O, summing both components.}}
  \end{flushleft}
	\label{tab:intens}
\end{table*}

For positions ND, SB, SC and SD, the H$_{2}$ column densities derived from the
dust and used to estimate the abundance of C$^{18}$O were calculated using a
dust temperature of 10K.  Assuming a temperature of 15K for all 4 positions
(ND, SB, SC and SD) would reduce the H$_{2}$ column densities by a factor of
2.1, representing an equivalent rise of the fractal abundance of C$^{18}$O by
the same amount.

The derived C$^{18}$O fractional abundance (which is averaged along the line
of sight) implies a depletion of C$^{18}$O of between a factor of 1.4 (for NC)
and 4.3 (for SC), with an average of 2.5 compared to the abundance of
$1.7\times10^{-7}$ in dark clouds \citep{1982ApJ...262..590F}.  Given that the
ratio between C$^{17}$O and C$^{18}$O has shown these two species to be
optically thin, with an intensity ratio of $\sim$3.5, a factor 2.5 depletion
of C$^{18}$O implies the same depletion factor for C$^{17}$O.


\end{appendix}

\end{document}